\documentclass[%
 reprint,
superscriptaddress,
preprintnumbers,
nofootinbib,
 amsmath,amssymb,
 aps,
pre,
]{revtex4-2}

\usepackage{amsmath,amssymb,amsfonts}
\usepackage{graphicx}
\usepackage{dcolumn}
\usepackage{multirow}
\usepackage{bm}
\usepackage[breaklinks=true,
colorlinks=true,
linkcolor=blue,
citecolor=blue,
urlcolor=blue]{hyperref}

\usepackage[english]{babel}
\usepackage{blindtext}
\usepackage{xcolor}

\usepackage{xcolor}

\usepackage[normalem]{ulem}

\begin{document}

\title{Predicting Atomistic Transitions with Transformers}

\author{Henry Tischler}
\thanks{Corresponding author}
\email{hentisch@lanl.gov}
\affiliation{%
Computing and Artificial Intelligence Division, 
Los Alamos National Laboratory, Los Alamos, NM 87545}
\affiliation{School of Engineering and Computer Science, University of Denver, Denver, CO 80210}
\affiliation{Department of Physics and Astronomy, University of Denver, Denver, CO 80210}

\author{Wenting Li}
\affiliation{Department of Electrical and Computer Engineering, University of Texas at Austin, Austin, TX 78712}

\author{Qi Tang}
\affiliation{School of Computational Science and Engineering, Georgia Institute of Technology, Atlanta, GA 30332}
\affiliation{Theoretical Division, Los Alamos National Laboratory, Los Alamos, NM 87545}

\author{Danny Perez}
\affiliation{X Computational Physics Division, Los Alamos National Laboratory, Los Alamos, NM 87545}

\author{Thomas Vogel}
\email{thomasvogel@lanl.gov}
\affiliation{Computing and Artificial Intelligence Division, Los Alamos National Laboratory, Los Alamos, NM 87545}

\date{April 12, 2026}

\begin{abstract}
Accurate knowledge of the atomistic transition pathways in materials and material surfaces is crucial for many material science problems. However, conventional simulation techniques used to find these transitions are extremely computationally intensive. Even with large-scale, accelerated material simulations, the computational cost constrains the applicable domain in practice. Machine learning models, with the potential to learn the complex emergent behaviors governing atomistic transitions as a fast surrogate model, have great promise to predict transitions with a vastly reduced computational cost. Here, we demonstrate how transformers can be trained to predict atomistic transitions in nano-clusters. We show how we evaluate physical validity of the predictions and how a~multitude of additional, different microstates can be generated by slightly varying the data provided to the model.
\end{abstract}

\maketitle

\section{Introduction}

Predicting the slow micro-structural evolution of materials is central to many problems in material science, for example, to predict the evolution of irradiated materials in fission or fusion applications~\cite{perez2014diffusion,domain2004simulation}. The separation of timescales between atomic vibrations, which control time steps in direct simulations based on molecular dynamics (MD) and ``interesting'' long-lived topological changes like defect nucleation, diffusion, or reactions, makes brute force approaches extremely computationally intensive \cite{uberuaga2020computational}. Instead, the evolution of systems through such sequences of rare events are often approximated by discretized ``state-to-state'' trajectories between different metastable states of the system. Such trajectories are considered to be statistically correct if the sequence of transitions is sampled from the physically correct probability distribution out of all possible transitions. These discrete trajectories are often generated using Kinetic Monte Carlo (KMC)~\cite{voter2007introduction} models
or (accelerated) MD~\cite{Perez2009ar,perez2024recent} simulations.
KMC models are computationally efficient but rely on extensive prior knowledge of all possible transition pathways and barriers. On the other hand, MD simulations do not rely on such prior knowledge but can be computationally expensive for systems with complex kinetics since transitions have to be generated dynamically, even in artificial, simulated conditions where they occur faster than in nature.

The direct simulation of the long-time evolution of materials using MD methods requires calculating interactions between all atoms and advancing atomic positions using extremely small time steps. Furthermore, the final states reached by transitions are often the result of larger emergent behaviors in the system, which can make it difficult to accurately simulate these behaviors using conventional algorithms. This observation motivated us to extend the recent advancement in generative AI to atomistic transitions and replace this process with a~generative model. By training a machine learning (ML) model on a large number of known transitions, we may learn these broader emergent long-time behaviors accurately and produce a surrogate model to directly predict transitions with significant savings in computational cost. By creating a tool that can quickly sample relevant final states, we can potentially make saddle configuration searches, that aim to enumerate a portion of all possible transitions, much quicker. Further, a probabilistic model could  extend this ability by sampling unit transition pathways and final state pairs with the correct physical probability, that is, proportional to the transition rate, conditioned on a given initial state. This would allow for statistically correct state-to-state trajectories to be predicted accurately, ultimately ``replacing'' traditional simulations entirely. 

We will focus here on a nano-cluster consisting of 147 platinum (Pt) atoms; 147 being the magic number for the Mackay icosahedron with three shells~\cite{mackay1962ac,schnabel2009cpl,Huang2017jcp}. This cluster has been studied previously using accelerated molecular dynamics (AMD) simulations~\cite{Huang2017jcp,Huang2018prm} and showed a variety of shape fluctuations and surface restructuring, which is typical for nanoparticles and can significantly affect their physical properties~\cite{baletto2005rmp}. This rich dynamical behavior makes these clusters ideal test environments to study the application of machine learning to predict atomic transitions. We use transitions found in AMD runs to train our ML models; the dataset consists of initial and final state pairs, represented by the positions of all atoms in three-dimensional space.

\begin{figure}
    \includegraphics[width=\columnwidth]{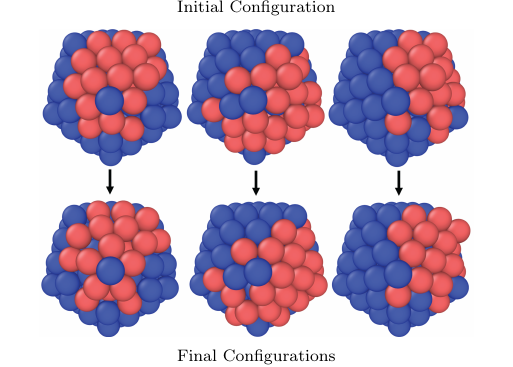}
    \caption{Three exemplar atomistic transitions originating from the same initial state of a 147-atom Pt nanocluster. Note that while all atoms are of the same element (Pt), we highlight atoms that are involved in each transition, that is, that significantly change their location during the transition, in red. Top row: The initial state, which is the same in each transition, but duplicated to show different colorings for the different atoms involved in each transition. Bottom row: Three different final states.}
    \label{fig:ex_transitions}
\end{figure}

One major challenge is the intrinsically stochastic nature of microstructure evolution in materials and atomistic transitions in particular. For every given atomistic configuration, there are often a large number of potential pathways to other nearby states the system can transition into, each with a different kinetic probability. See Fig.~\ref{fig:ex_transitions} for some examples. Furthermore, such pathways can be highly complex and collective, leading to a combinatorial explosion of the complexity of the problem.

Machine-learning based surrogates have been proposed as a way to predict long-time dynamics in physics systems. A notable successful example is longer-time global weather prediction~\cite{lam2023learning}. Similarly to the proposed work, flow maps are often learned for long-time prediction, using, for example, symplectic neural networks as a fast surrogate to generate Poincare plots~\cite{burby2020fast}. Physics-constrained neural networks have also been trained to generate long-time dynamics of collisional radiative models~\cite{xie2024latent}. An alternative approach are neural ordinary differential equations~\cite{chen2018neural} that learn the right-hand side of the latent dynamics as a surrogate model; applied examples include the long-time dynamics in singularly perturbed dynamical systems~\cite{serino2025fast} and stiff dynamics such as chemical reactions~\cite{caldana2024neural, loya2025structure}.

In a seminal work~\cite{vaswani2023attentionneed}, the transformer architecture was proposed for natural language processing, which can naturally be extended to dynamical predictions. Transformers have indeed since been used to predict  shock dynamics in inertial confinement fusion ~\cite{serino2024reconstructing} and are the main focus of the current study, where we consider small metallic nano-clusters~\cite{Huang2017jcp,Huang2018prm} as a prototype to explore the prediction power of machine learning surrogates for predicting atomistic transitions.
We here report a proof-of-concept demonstration of a first step towards predicting atomistic transitions using transformer architectures. Specifically, we implemented and trained a transformer model to predict one of the many possible transition paths for a given initial configuration.
We then preconditioned the model with what we call ``hints'' that provide additional information about the final state geometry, allowing our model to deterministically distinguish a specific final state from other physically plausible configurations. 
We demonstrate that our model can often predict specific transitions even when very minimal information about the final state is provided. Moreover, our model can consistently predict many different physically plausible transitions without this hint.

\section{Methods}
To produce a proof of concept for predicting atomistic transitions using transformers, we first produced a large dataset of pairs of initial and final states in such transitions (see Fig.~\ref{fig:ex_transitions} again for examples) using direct, accelerated molecular dynamics simulations. We then train a transformer model to predict final states from each initial state.
A major challenge in evaluating this model arises from the fact that such systems exhibit multiple, complex transition pathways for each initial state. Although we can discover some of these transitions from simulation, enumerating all possible final states is computationally infeasible. In order to first verify that our model can predict transitions that are already known, we guide the model towards one specific, out of the many possible, final states by providing what we call a ``hint'': additional information about the known final state we expect the model to generate. We then test if the transformer can predict the movements of the remaining atoms in the transition. 

We also investigate the performance of the model when decreasing the hint size and show that our model can predict previously unknown final states. We verify these newly discovered transitions using an explicit Nudged Elastic Band (NEB) calculation~\cite{jonsson1998neb}, to confirm that these are proper single-step transitions that are physically plausible. Most importantly, we find that even without providing any hints whatsoever to the transformer, it is able to predict new, physically relevant transitions. In this section, we describe in detail how we prepare the dataset, set up the transformer, and provide different types of ``hints''. We also discuss how we identify equivalent states and transitions and how we check the physical validity of previously unknown transitions. 

\subsection{Dataset}
For model training and validation, we use a dataset of 239,594 transitions between 86,543 low-local energy minima in a platinum nano-cluster containing 147 atoms. These trajectories were produced from an accelerated MD simulation carried out using the Parallel Trajectory Splicing (ParSplice) method~\cite{perez2016long}. ParSplice uses direct MD simulations as a transition discovery engine in a way that is highly accurate and provably unbiased. 
The interactions between atoms were represented using an embedded-atom model potential~\cite{voter1993}. The dataset includes physical positions in three dimensional Cartesian space for each atom.

Before training, we preprocessed the initial and final states of each transition to remove global rotations, reflections, and translations using a standard least-squares alignment algorithm. In addition, we ordered the atoms in each configuration by their distance from the configuration's center of mass.\footnote{This sorting is performed for the initial state and then applied to the final state, in order to provide consistent ordering between these configurations.} As these characteristics are matched during data preprocessing, the transformer model avoids having to learn these characteristics as equivariant. 

We split our dataset into two sub-sets for training and validation such that none of the configurations in the validation dataset could be seen by the model during training. This allows us to establish that our model can generalize to configurations with explicitly unseen geometries, rather than overfitting to previously seen configurations. Our training dataset contained 198,797 transitions between 74,260 configurations, while the validation dataset contained 40,797 transitions between 12,283 configurations. This resulted in an approximate 82/18 split between our training and test dataset in terms of \hbox{transitions}. 

\subsection{Model Architecture}

\begin{figure}
    \centering
    \includegraphics[width=\linewidth]{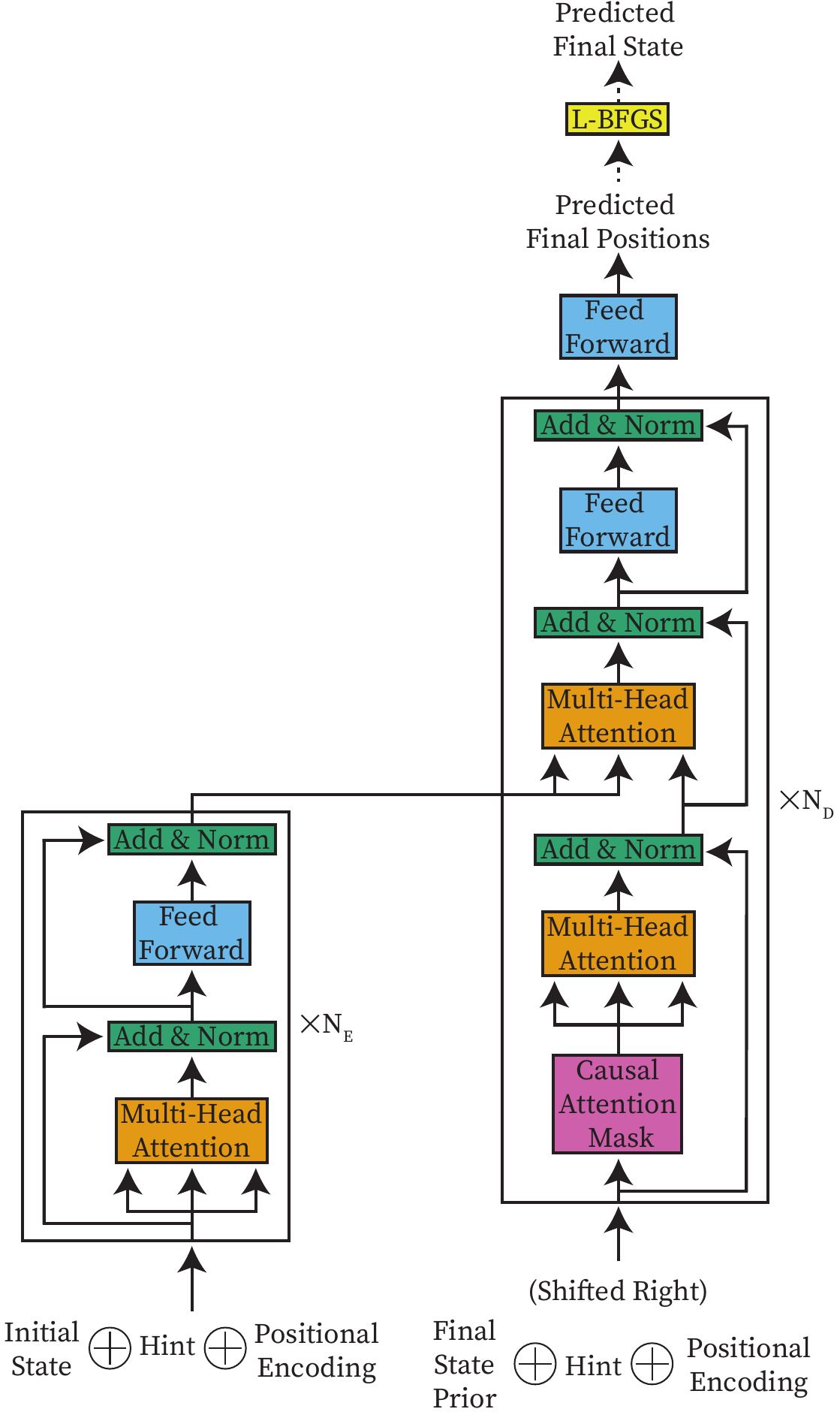}
    \caption{Diagram of our transformer model. The architecture is nearly identical to the one introduced in~\cite{vaswani2023attentionneed}, though the input encoding scheme is modified to allow continuous input.  Dotted lines indicate steps not involved in training.}
    \label{fig:transformer-setup}
\end{figure}

Our architecture is nearly identical to the original transformer proposed in~\cite{vaswani2023attentionneed}. However, the input encoding scheme is modified to allow for the representation of continuous atomistic configurations, see Fig.~\ref{fig:transformer-setup}. Our model therefore avoids the typical tokenization step used for training transformers on discrete objects (for example, words). Instead, we directly replace the embedding which would normally be learned for a particular token by simply concatenating the atomic position and a positional encoding:
\begin{equation}
\mathbf{X}_i = \left(x_i, y_i, z_i, p_{\{i,1\}}, \ldots, p_{\{i,n\}}\right)
\end{equation}
where $x_i$, $y_i$, and $z_i$ are the Cartesian coordinates and $\vec{p}_i$ provides the positional encoding for the $i$th atom. $\mathbf{X}_i$ is then the encoded vector given to the first multi-head attention blocks in our model for each atom. We use a positional encoding~\cite{vaswani2023attentionneed} of length $n=157$, such that each atomic position is represented by an input vector of size 160. We found this to be an effective choice and were unable to achieve similar performance with positional encodings substantially shorter.

During the prediction stage, we use an Limited-Memory Broyden–Flet\-cher–Goldfarb–Shanno (L-BFGS) optimizer~\cite{Liu1989} to minimize the predicted final positions closer to their nearest metastable configuration within the potential energy landscape. We allow 10,000 steps to converge with a force tolerance of 0.0001 eV/\AA. Other parameters were not changed from the default values provided by the ASE library~\cite{hjorth1017ase-paper}.

\subsection{Identifying Identical Transitions}
\begin{figure}[b!]
    \centering
    \includegraphics[width=.8\columnwidth]{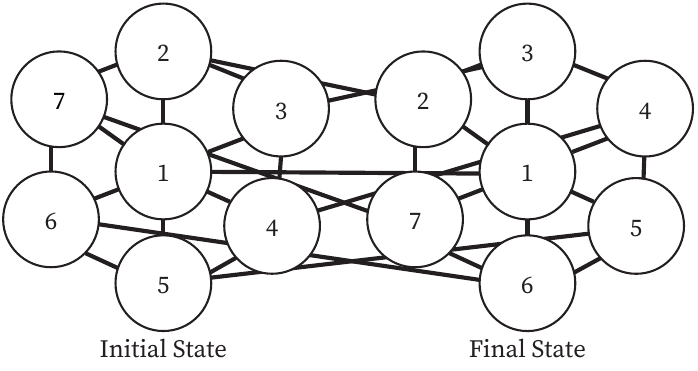}
    \caption{A bi-partite connectivity graph for an illustrative movement in a cluster of 7 atoms. Atom indices give arbitrary indexings.}
    \label{fig:bipartite-ex}
\end{figure}

To test our models, it is very useful to assess whether our predicted and known transitions are physically equivalent. We define transitions as equivalent if they contain the same set of nearest-neighbor relationships.
In practice, we use a method proposed in~\cite{beland2011kinetic} to construct connectivity graphs for each initial and final state, which give each nearest-neighbor relationship as an edge on this graph. We then combine these into a bipartite connectivity graph where an edge is drawn between each atoms representation in the initial and final states. This allows us to distinguish transitions which are distinct in the atomic movements produced, but produce symmetrical final states that would otherwise produce isomorphic connectivity graphs. An example of such a bipartite graph is shown in Fig.~\ref{fig:bipartite-ex}. We classify transitions as equivalent if their bi-partite connectivity graphs are isomorphic when constructed from our known and predicted transitions.
\begin{figure}[b!]
    \centering
    \includegraphics[width=.95\columnwidth]{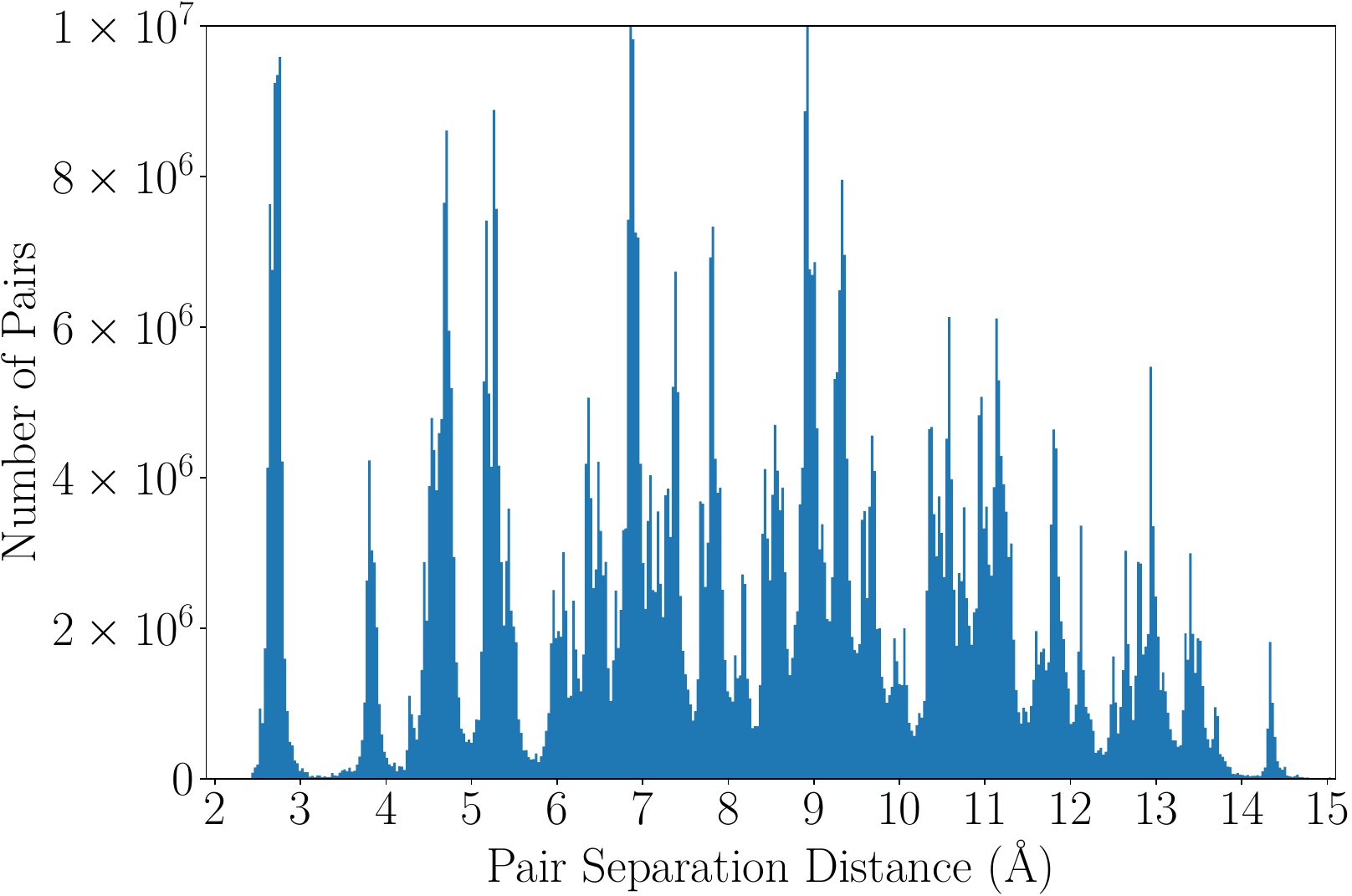}
    \caption{The cumulative radial distribution function, which aggregates the pair separation distances across all configurations. The position of the valley separating the first and second peak, 3.2\,{\AA}, is chosen as the connectivity threshold for creating a connectivity graph for each Pt cluster.}
    \label{fig:rdf}\vspace{-1.2mm}
\end{figure}

In the fcc lattice, platinum has a lattice constant of 3.9\,{\AA}, resulting in a distance of approximately 2.8\,{\AA} between Pt atoms in a crystal. This distance corresponds to the position of the first peak in a radial distribution function, as shown in Fig.~\ref{fig:rdf}. Due to the the presence of different phases and of elastic distortions, this peak has some width. We use a distance of 3.2\,{\AA} between atoms as the connectivity threshold to reliably capture all nearest neighbor pairs in a cluster. In Fig.~\ref{fig:rdf}, this distance corresponds to the position of the valley between the first two peaks, separating nearest and next-nearest neighbors.

\subsection{Identifying Relevant Transitions}

Because of the very high dimensionality of the potential energy surface of the system and the complexity of the configurations that were generated, we should expect that each initial state can experience a high number of transitions to different final states, most of which cannot be expected to be present in the training set. To evaluate the validity of such previously-unseen transitions, we reconstruct the minimum energy pathway (MEP) between the input and output state using a nudged elastic band (NEB) test~\cite{jonsson1998neb}. Based on this reconstructed pathway, we here consider a transition as ``physical'', or dynamically ``relevant'', if it has a single, reasonably small energy barrier (and, thus, a reasonably high transition rate). We approximate a transition rate for these transitions using the vineyard formula~\cite{vineyard1957}, using a so-called ``standard'' vibrational prefactor of $10^{-12}\,1/\mathrm{s}$ which roughly corresponds to a typical vibrational frequency in solids. Transitions should be called ``unphysical'' if the MEP does not show a barrier or has multiple barriers, and dynamically ``irrelevant'' if the barrier is so high that transition times become astronomical. We will use the notion of a ``correct'' prediction for the case where we hint the transformer with the goal of reproducing a specific final state contained in the simulation dataset.

We note that although all energy pathways presented below appear to be ``relevant'', this determination is, for now, made from visual inspection of the energy pathway. Thus, the accuracy of the presented transition rates is not guaranteed beyond this approximation.

\section{Training}
As is often the case for large generative models, our transformers require significant compute resources for training. Here we layout the practical resources and parameters used to train our models and how these parameters are selected. 
Although the process used to train both versions of our transformer model are identical, the specific parameters used for training are different.

\subsection{Computing Resources}
Both variations of our model were trained on a compute node with 8 NVIDIA Titan V GPU's. The version of our model provided with magnitude hints was trained for close to two hours, and the version of our model provided with partial-position hints was trained for less than an hour. The difference in training time is a result of the difference in model sizes.

\subsection{Hyperparameter Scan}
For both variations of our model, we conducted a large-scale, randomized hyperparameter search to find an effective set of hyperparameters.
The ranges of the hyperparameters are presented in Table~\ref{tab:hyperparameter-space}.
\begin{table}[h!]
\centering
\caption{The search space for hyperparameter scans. Hyperparameters that vary continuously were sampled uniformly.\vspace{.5em}}\label{table:hyperparameters}
\begin{tabular*}{\columnwidth}{@{\extracolsep{\fill}}|ll|}
\hline
Hyperparameter & Search Space \\ \hline
&\\[-.25cm]
Learning Rate & {[}0.0001, 0.1{]} \\
Number of Encoder Layers & \{1, 2, 3\} \\
Number of Decoder Layers & \{1, 2, 3\} \\
Number of Multi-Attention Heads & \{2, 4, 5, 10, 20\} \\
Batch Size & {[}16, 4096{]} \\
Feed Forward (FF) Hidden-Layer Size & \{1024, 2048\} \\ \hline
\end{tabular*}
\label{tab:hyperparameter-space}
\end{table}
\begin{table}[b!]
\caption{Optimized hyperparameter values for the selected ``partial-position-hinted'' model (PPH) and ``individual displacement'' model (IDM). See Table~\ref{tab:hyperparameter-space} for the search space for each parameter.\vspace{0.5em}}
\begin{tabular*}{\columnwidth}{@{\extracolsep{\fill}}|ccc|}
\hline
\multirow{2}{*}{Hyperparameter} & \multicolumn{2}{c|}{Value} \\
& PPH & IDM \\ \hline
&&\\[-.25cm]
Learning Rate & 0.0248 & 0.0777 \\
Number of Encoder Layers & 1 & 3 \\
Number of Decoder Layers & 2 & 2 \\
Number of Multi-Attention Heads & 10 & 10 \\
Batch Size & 389 & 197 \\
Transformer FF Hidden-Layer Size & 2048 & 2048 \\
Final Feed Forward Hidden-Layer Size & 2048 & 2048 \\ \hline
\end{tabular*}
\label{tab:hyperparameters}
\end{table}

\begin{figure*}
    \centering
    \includegraphics[width=\linewidth]{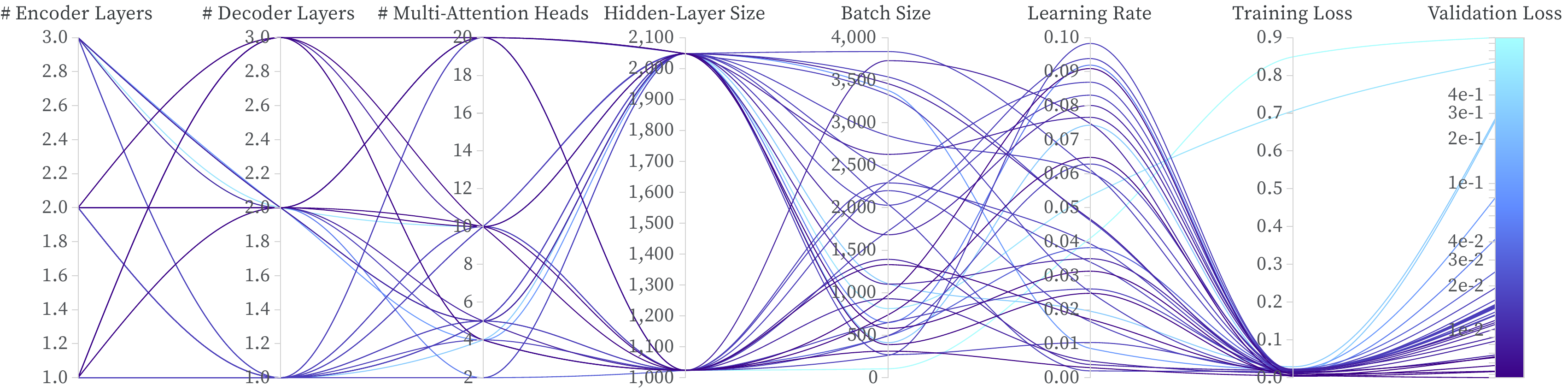}
    \includegraphics[width=\linewidth]{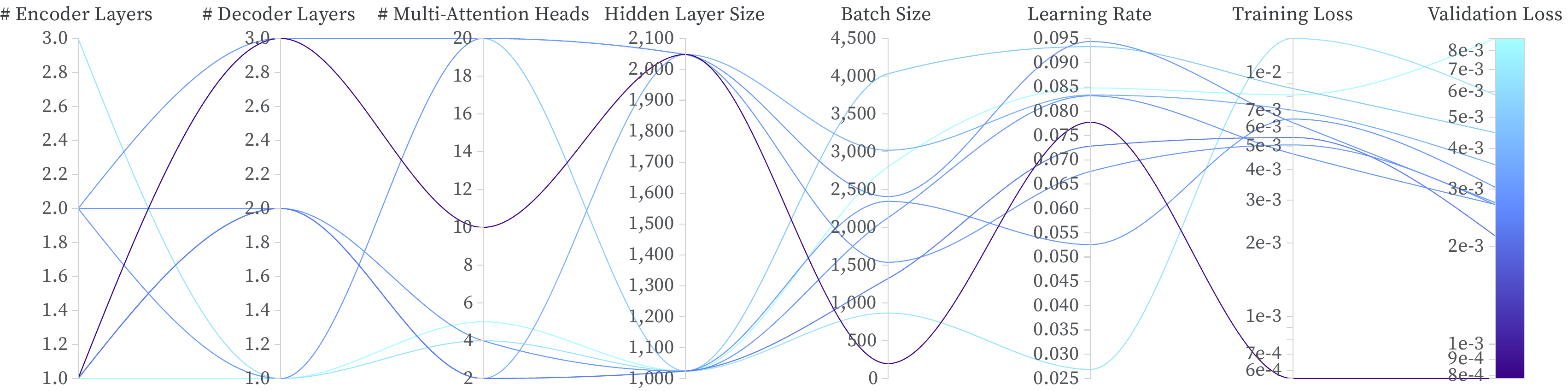}
    \caption{Losses from individual trials in our hyperparameter scan. Top: Results our model not provided with hints during training (partial-position hinted). Bottom: Results for our model provided with hints during training (individual magnitude hinted). Plots were generated using Weights \& Biases~\cite{wandb}.}
    \label{fig:sweep-plost}
\end{figure*}

Both hyperparameter scans were conducted across two compute nodes with 8 NVIDIA Titian V GPUs. For the individual-displacement hinted model, we tested ten randomized variations of hyperparameters, using approximately 10 hours of cumulative compute time across both nodes. For the model hinted with partial-final states, we tested 40 randomized variations of hyperparameters, using approximately 24 hours of cumulative compute time. These scans were automated using Weights and Biases~\cite{wandb}.

The per-trial results of these hyperparameter scans are shown in Fig. \ref{fig:sweep-plost}. We selected each final model by picking the version with the lowest validation loss.

\subsection{Model and Training Hyperparameters}
For all models, we used an Adam optimizer~\cite{adam} with $\beta_1 = 0.9$, $\beta_1 = 0.98$, and $\epsilon = 10^{-9}$. Each model was trained for 150 epochs and the learning rate was scheduled using the original scheduling scheme for the transformer~\cite{vaswani2023attentionneed}. The final model hyperparameters are given in Table~\ref{tab:hyperparameters}.

\section{Results}

Microstructure evolution in materials is inherently stochastic and a model should reflect that in a reasonable and useful way. In nature, for any given state of a~system, there is, in principle, a large number of possible other states the system can transition into with a certain probability. We are aiming at efficiently providing a broad spectrum of potential final states for a given input, that is, our model should predict a set of dynamically relevant states that have not been seen in simulations before. As a necessary first step however, we show that our model is able to correctly and reliably reproduce states that we do already know. For this, we provide the model with hints that will allow it to choose a particular state from an arbitrarily large number of potential final states. We will refer to these as ``hinted'' models (Sects~\ref{subsec:res_position_hint} and~\ref{subsec:res_displacement_hint}). To show that the model can also predict previously unknown, relevant states, we then reduce the magnitude of the hints to zero, resulting in the ``unhinted'' model limit. To further trigger the prediction of a set of different final states without providing information for those states, which one would not have for unknown states, we add a certain, small amount of random noise to the input state (Sect.~\ref{subsec:res_unhinted}).

\subsection{Partial Position Hinted Model}
\label{subsec:res_position_hint}

\begin{figure}
    \centering
    \includegraphics[width=.9\columnwidth]{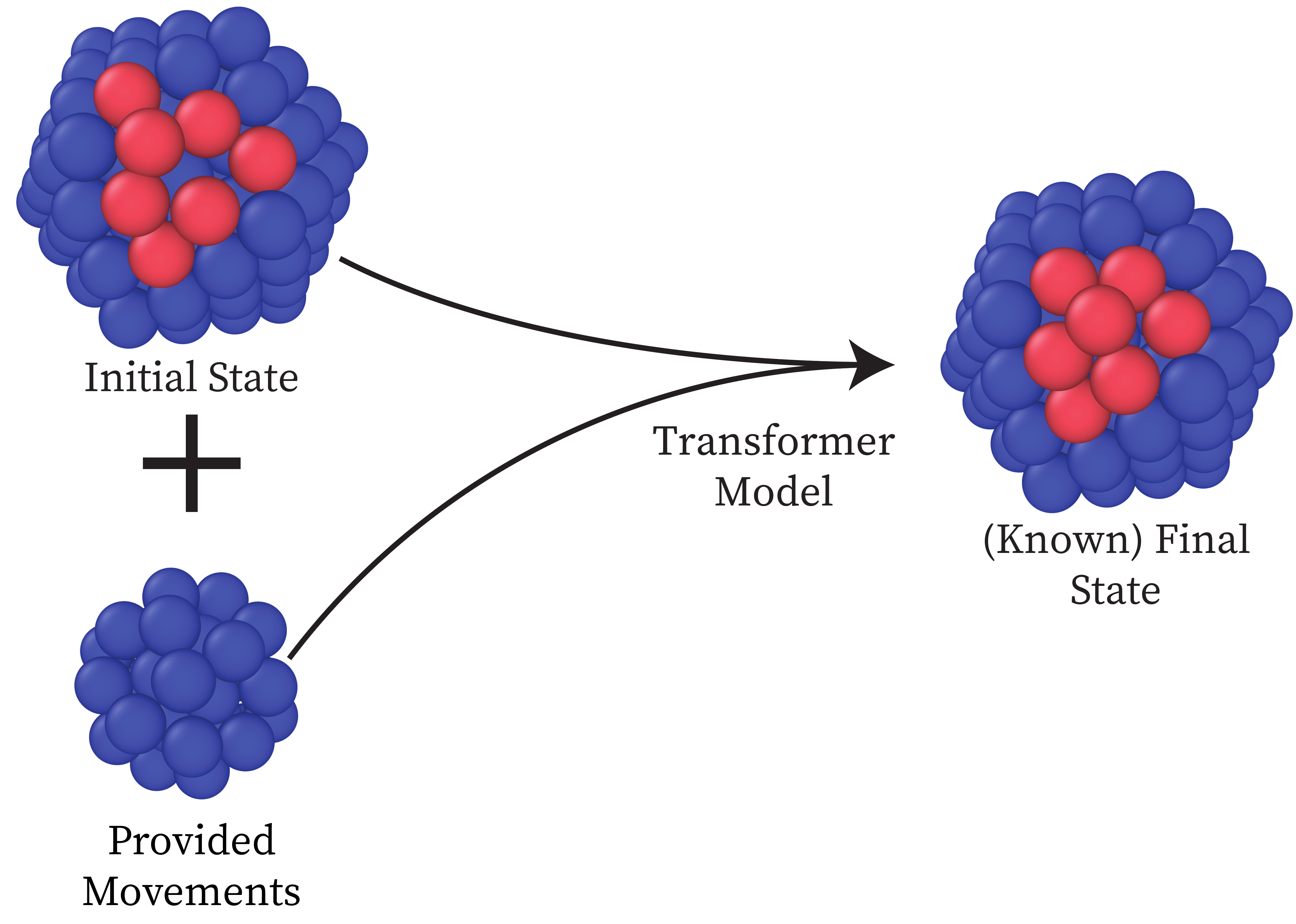}
    \caption{A diagram of our partial-position hinted model. Red colored atoms are those that change their position most during the transition. Top left: an initial state. Bottom left: positions of atoms in its final state that are provided to our model as a ``hint''. Right: the final state that is correctly predicted from the initial state and ``hint''.}
    \label{fig:partial-position-hint-diagram}
\end{figure}

In order to guide our model to predict final states that we know to exist without modifying the transformer architecture, we provide our model with what we call ``partial-position hints''. As illustrated in Fig.~\ref{fig:partial-position-hint-diagram}, these hints provide our model with some of the atomic positions in the final state. We then task our model with predicting the rest of the transition. The hints are provided directly as context to the transformer in its decoder prior (illustrated as the ``Final State Prior'' in Fig.~\ref{fig:transformer-setup}), which is otherwise used in an auto-regressive manner to precondition the model on the atoms it as already predicted during inference~\cite{vaswani2023attentionneed}. In order to match the format of the prior seen during training, the atoms in our hint are ordered from the center of mass outwards. 

To quantify how aggressive our hint is, we define a heuristic ``hint size'', $s_h$, as the ratio of the sum of the displacements of atoms that are given in the hint compared to the total displacement of all atoms under the target transition: 
\begin{equation}
s_h=\frac{ \sum_{i\in h}\lvert \vec{d}_i \rvert}{\sum_{i=1}^N \lvert \vec{d}_i \rvert}\label{eqn:metric}
\end{equation}
where $h$ is the set of atoms given in the hint, $N$ is the total number of atoms, and $\vec{d}$ is the displacement vector of each atom in the transition between the initial and final state.
Because we quantify the ``significance'' of a final state position in the hint exclusively from the magnitude of its movement between the initial and final states, there is no need to directly classify which atoms are involved in the transition, which can be difficult to identify in our physical system. However, our implicit assumption is that the amount of information provided in a single atomic position is directly proportional to its displacement during the transition is a limited approximation; for example, when significant information is given to the model by indicating that specific atoms do not transit, our metric will fail to capture this. 
In practice, however, we believe that the proposed metric (Eq.~\ref{eqn:metric}) is a reasonable heuristic measure for the amount of information provided to the transformer, especially in extreme cases when either none or all of the information about the final state is provided. As a more conservative way to estimate the amount of information provided in a hint, we also analyze the total number of positions included in this hint. This more directly measures the amount of additional information provided to our model, though likely overestimates how ``significant'' a hint is when many atoms do not provide our model with any useful information about the known final state.

\begin{figure}
\centering
\begin{tabular}{ccc}
Initial State & Partial-Position Hint & Final State \\[3mm]
\includegraphics[width=0.25\linewidth]{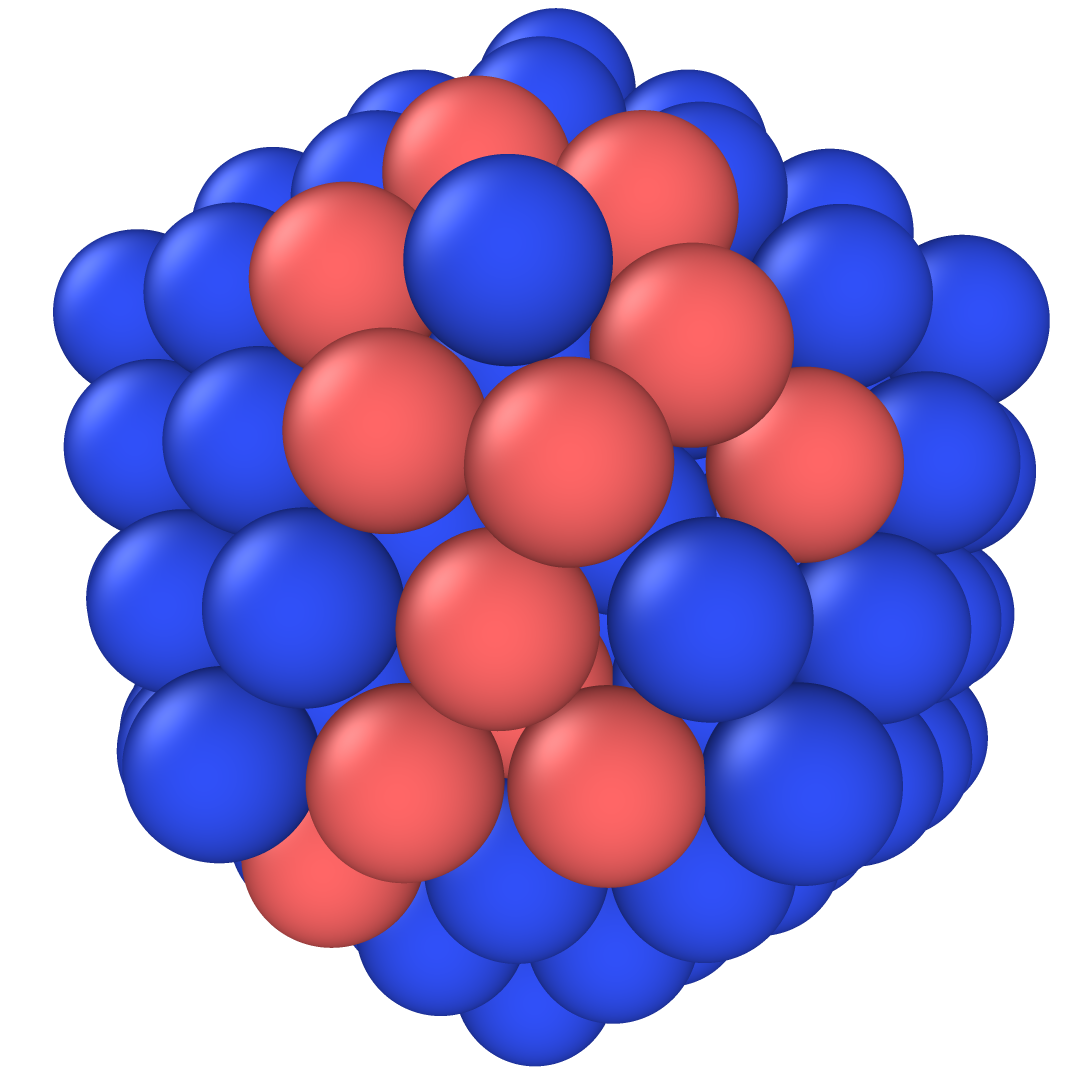} &
\includegraphics[width=0.25\linewidth]{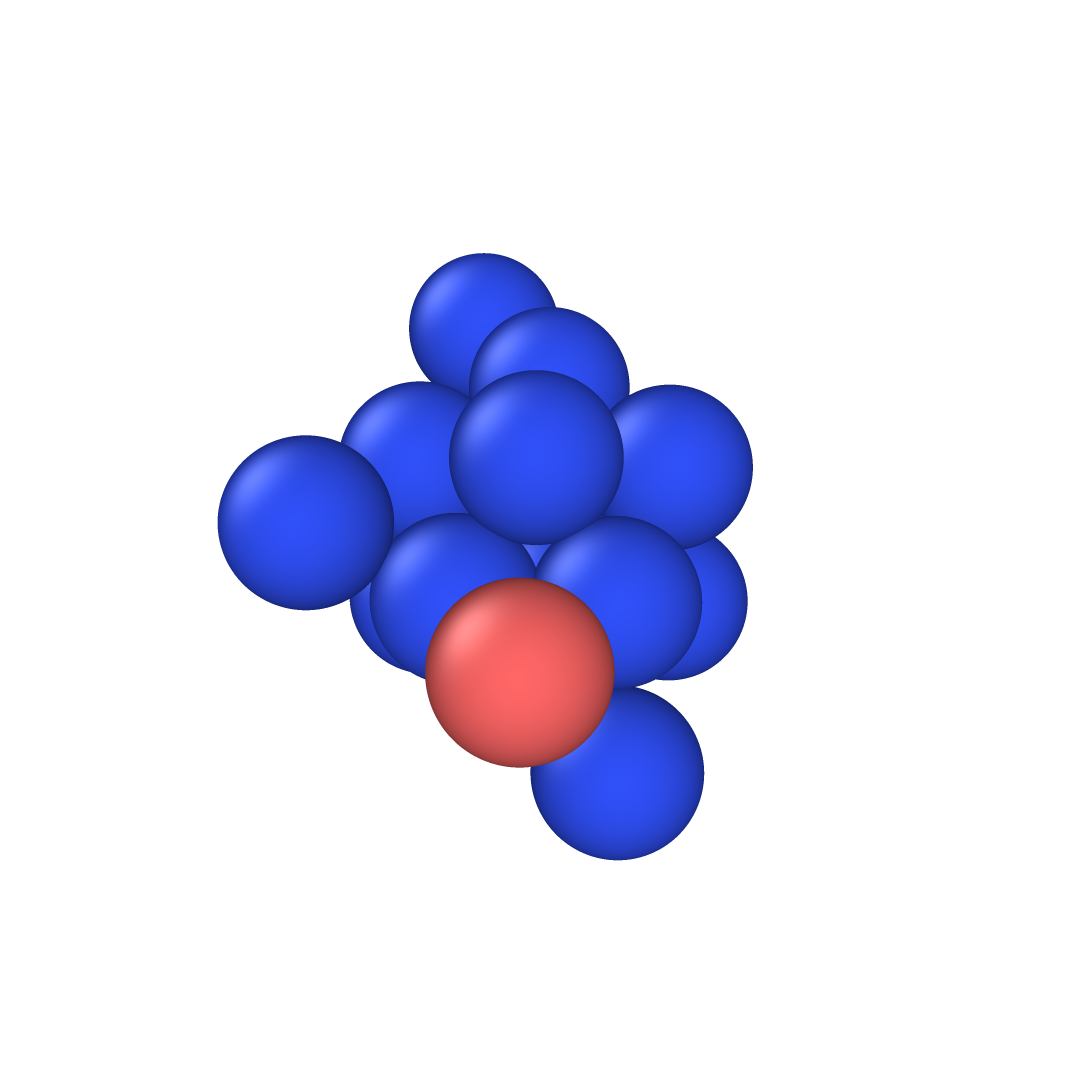} &
\includegraphics[width=0.25\linewidth]{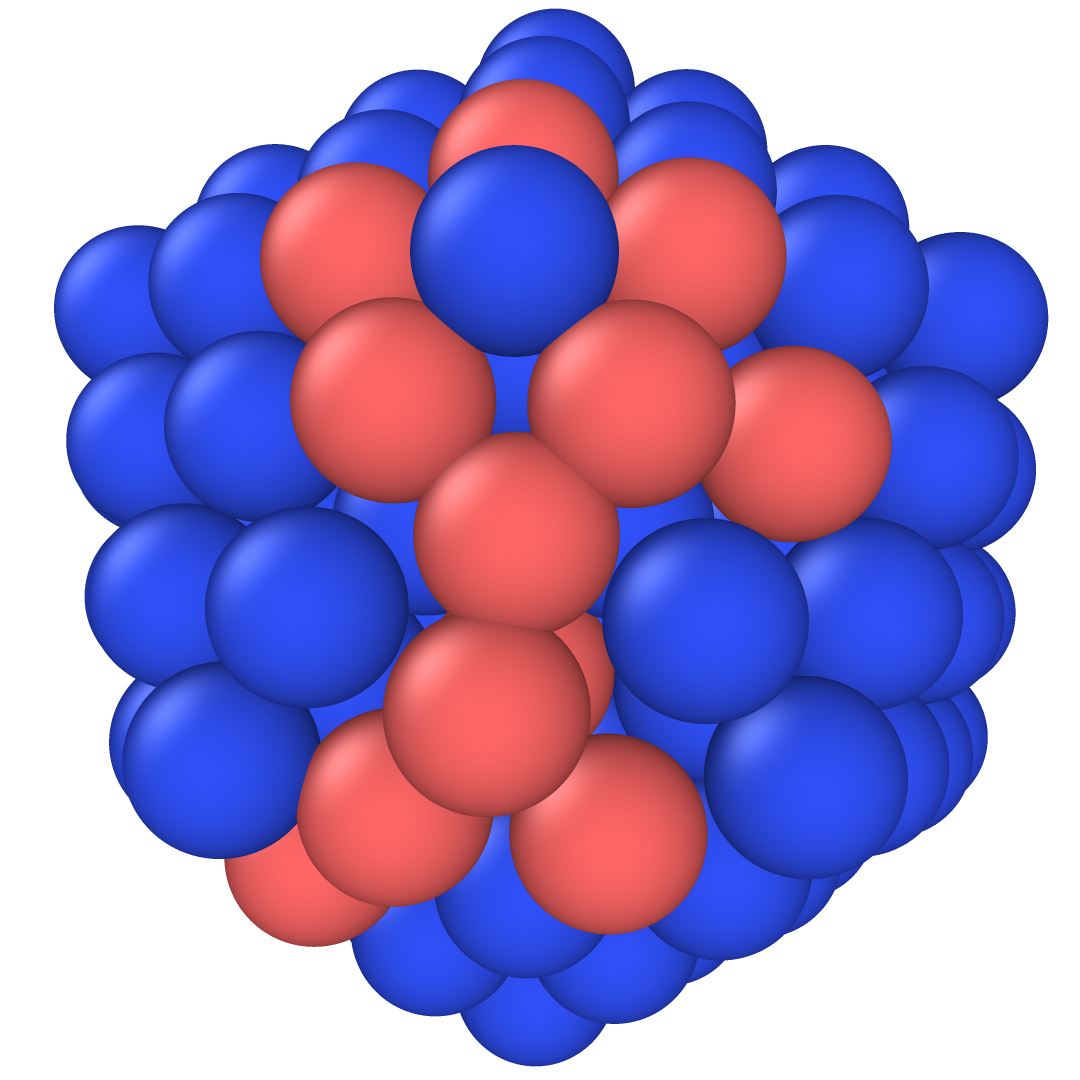} \\[2mm]
\includegraphics[width=0.25\linewidth]{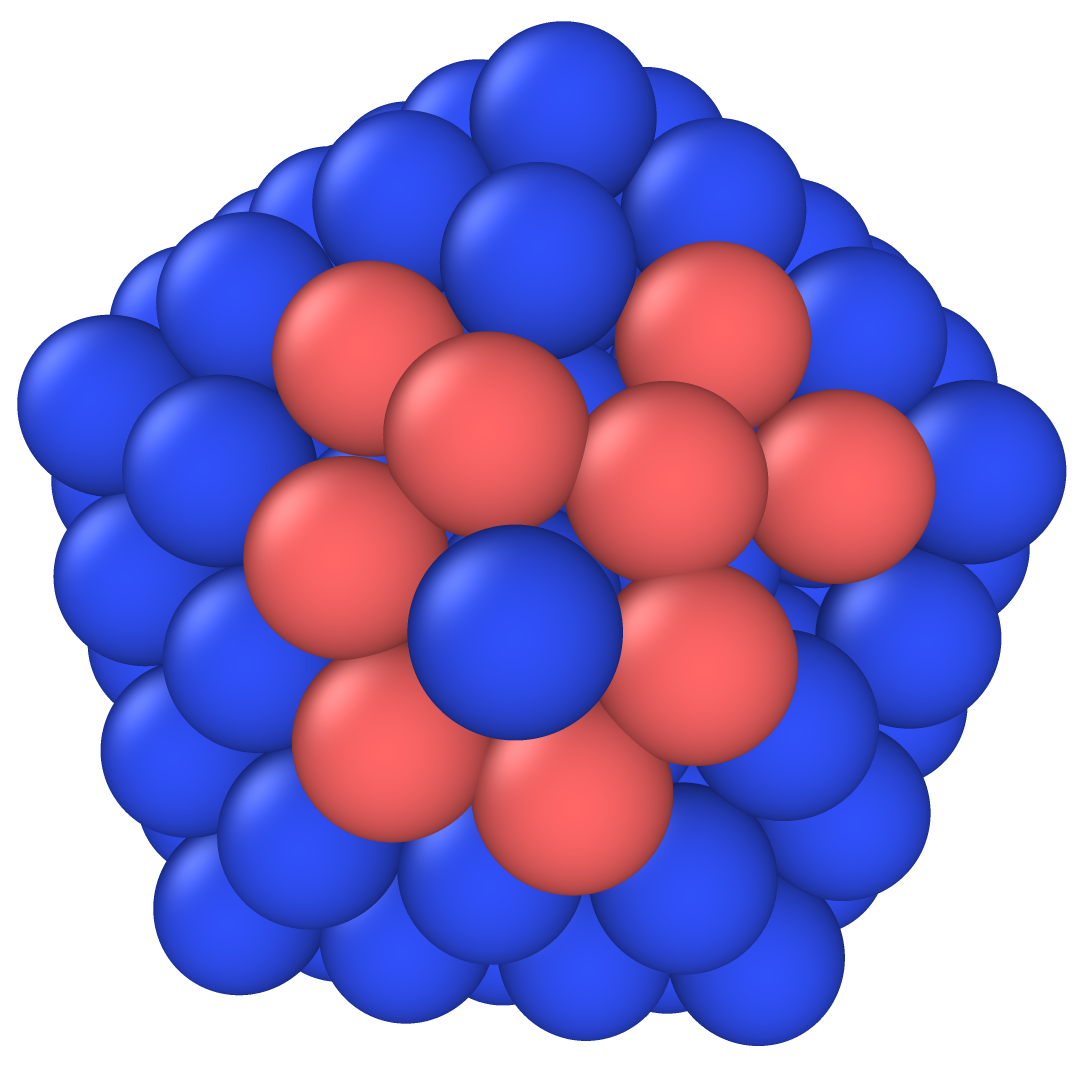} &
\includegraphics[width=0.25\linewidth]{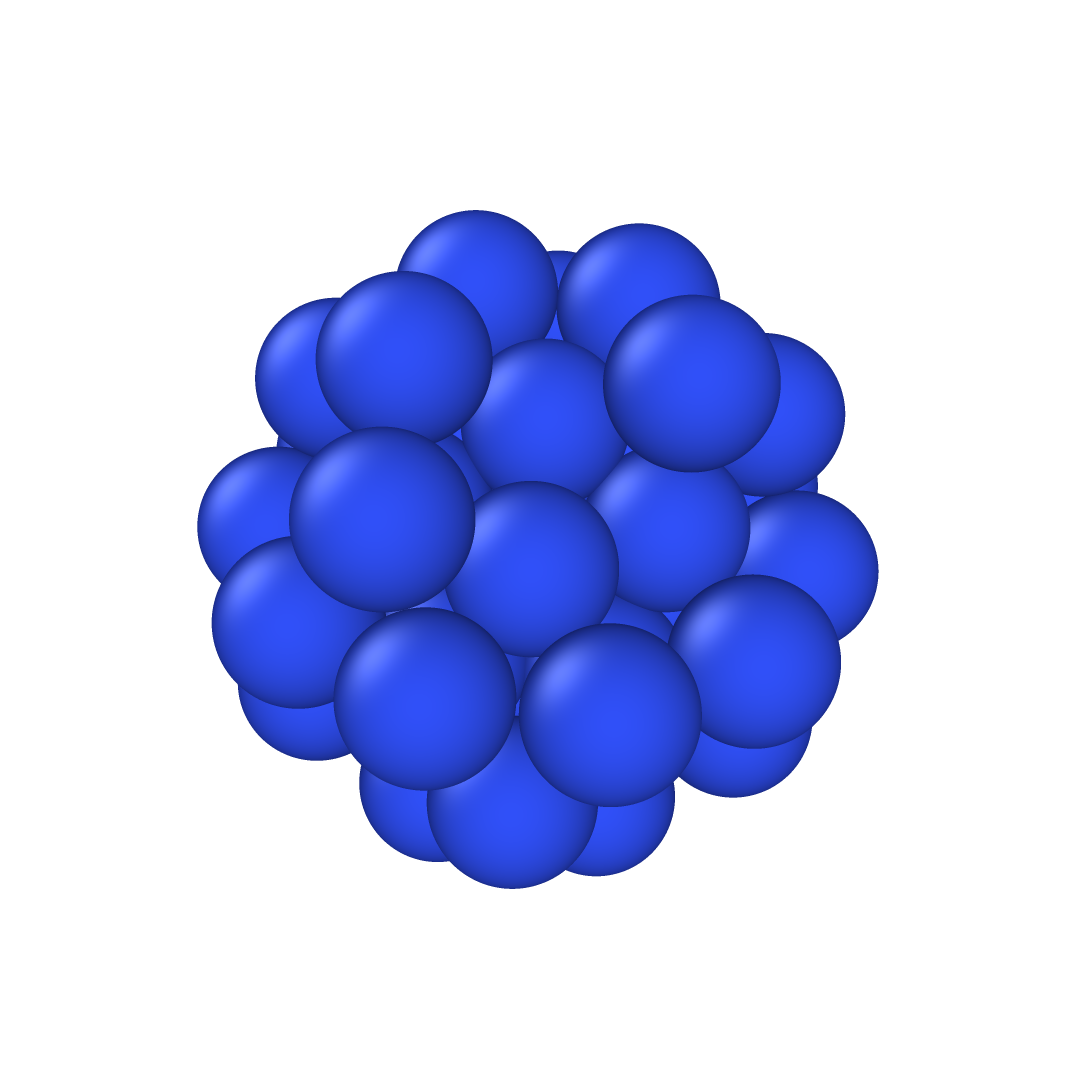} &
\includegraphics[width=0.25\linewidth]{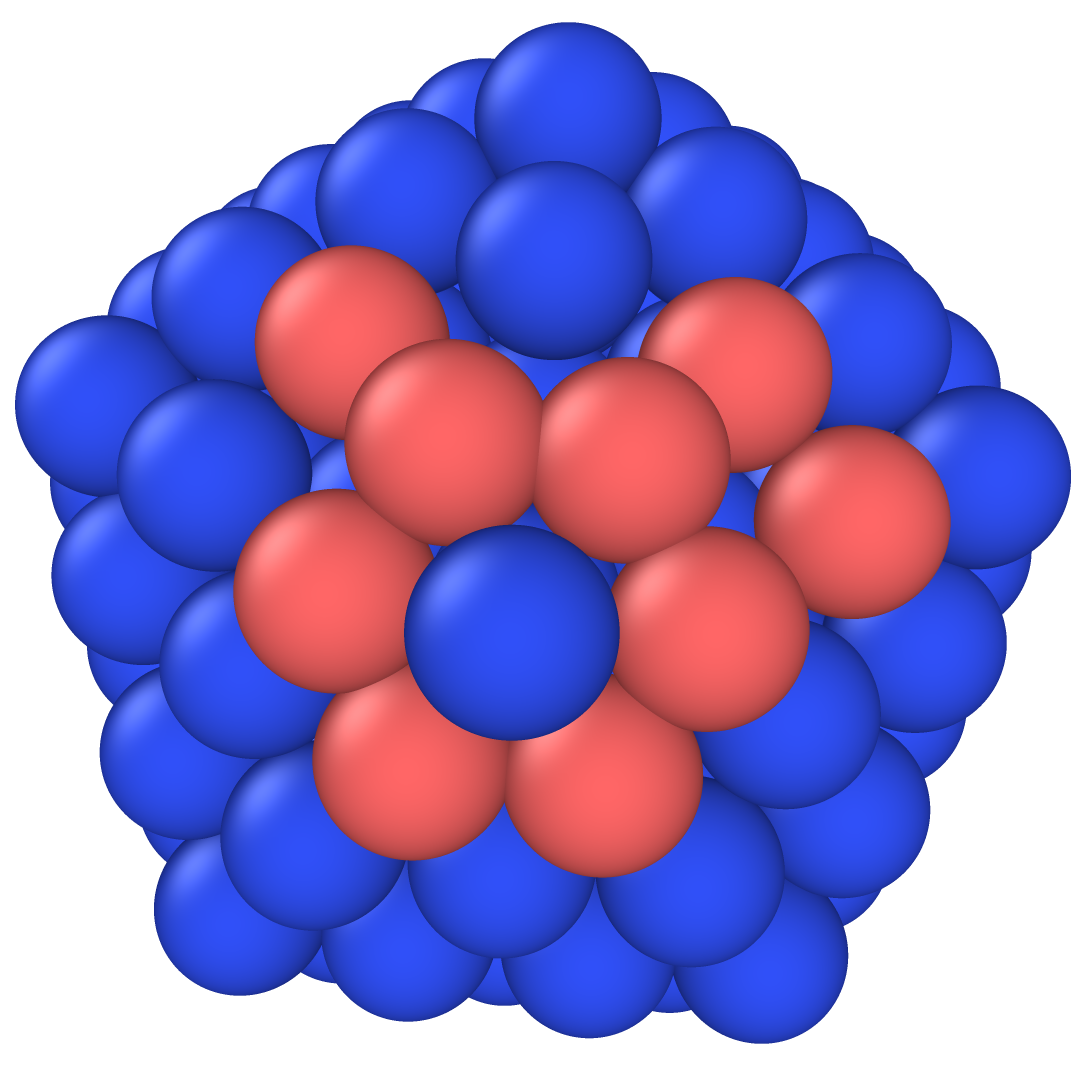} \\[2mm]
\includegraphics[width=0.25\linewidth]{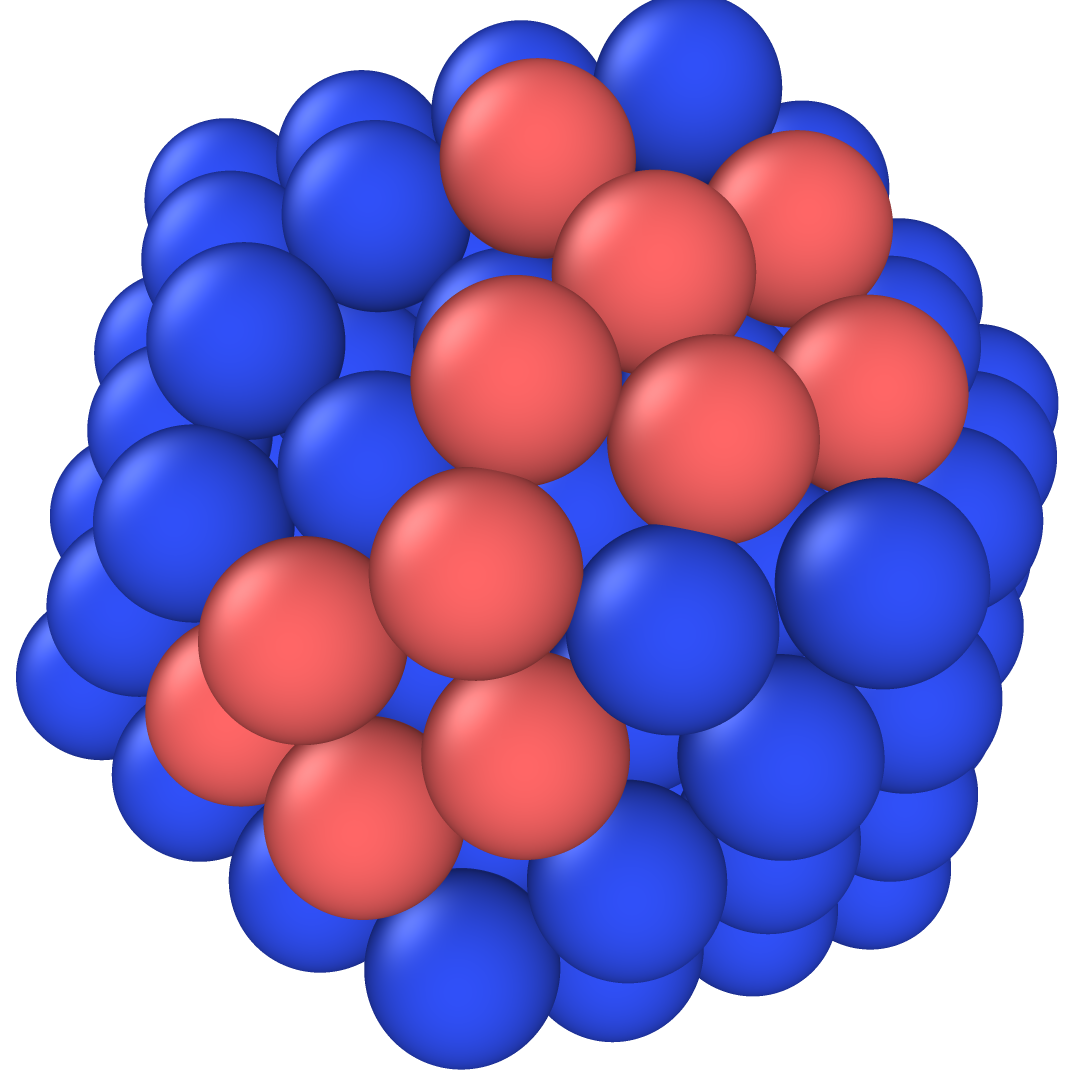} &
\includegraphics[width=0.25\linewidth]{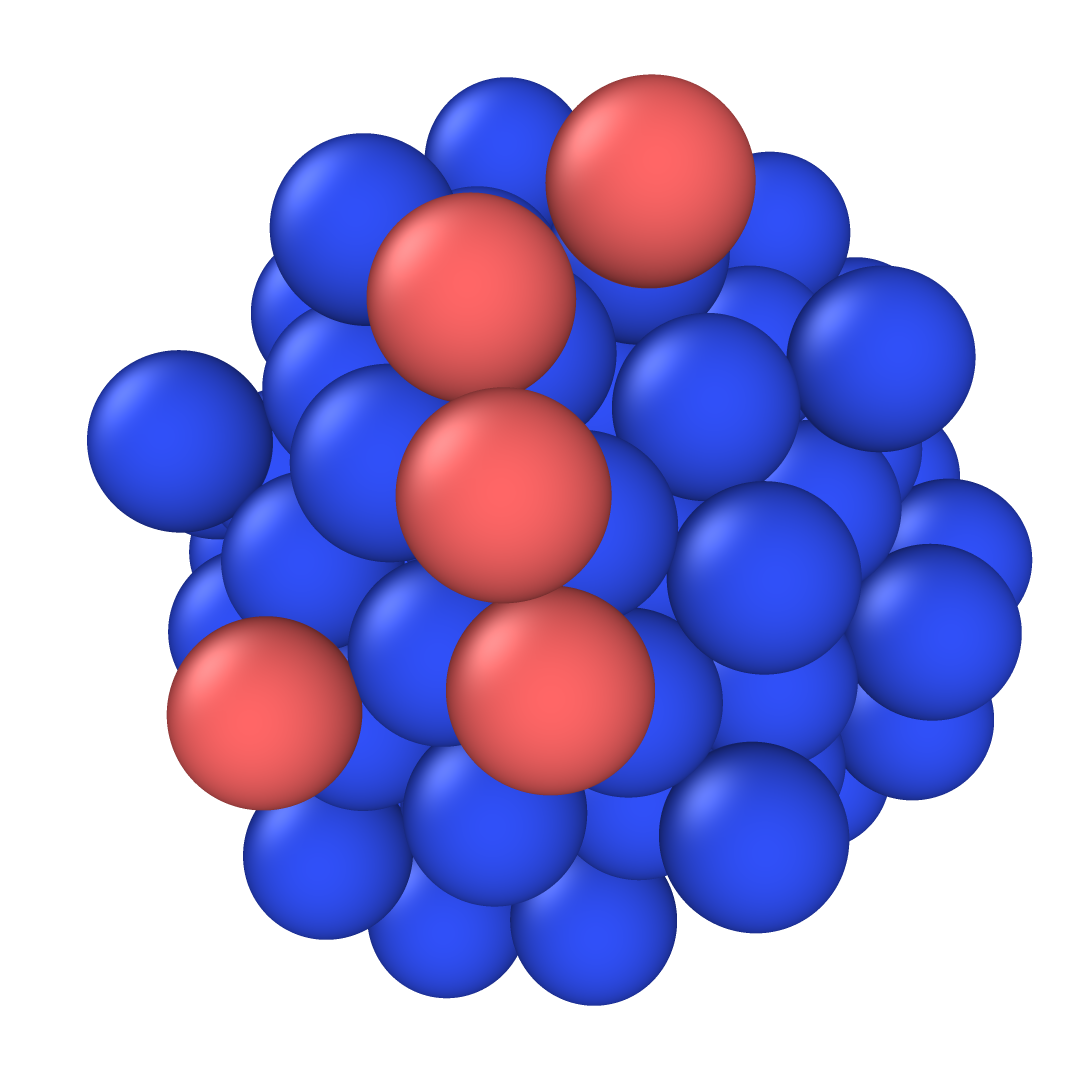} &
\includegraphics[width=0.25\linewidth]{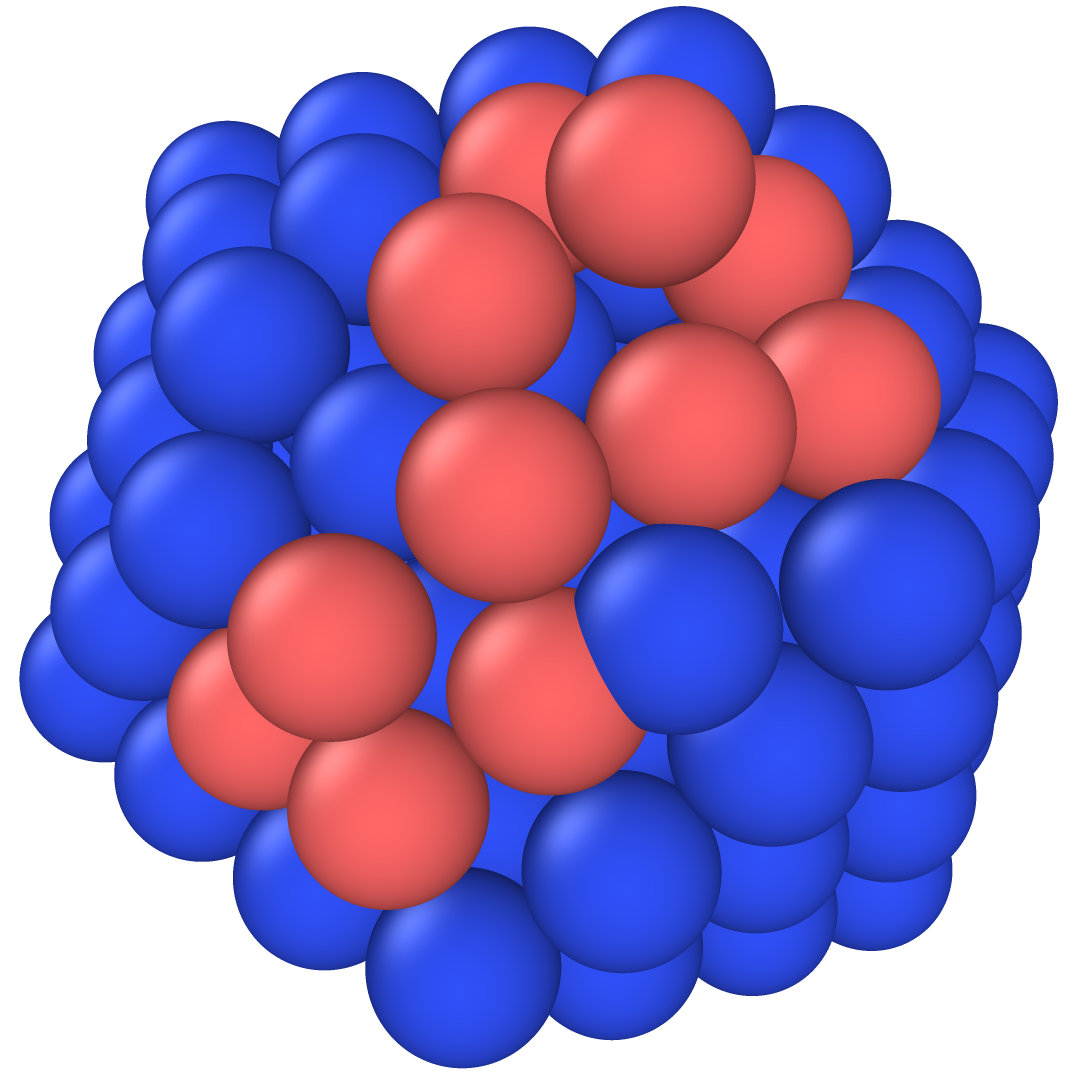} \\[2mm]
\includegraphics[width=0.25\linewidth]{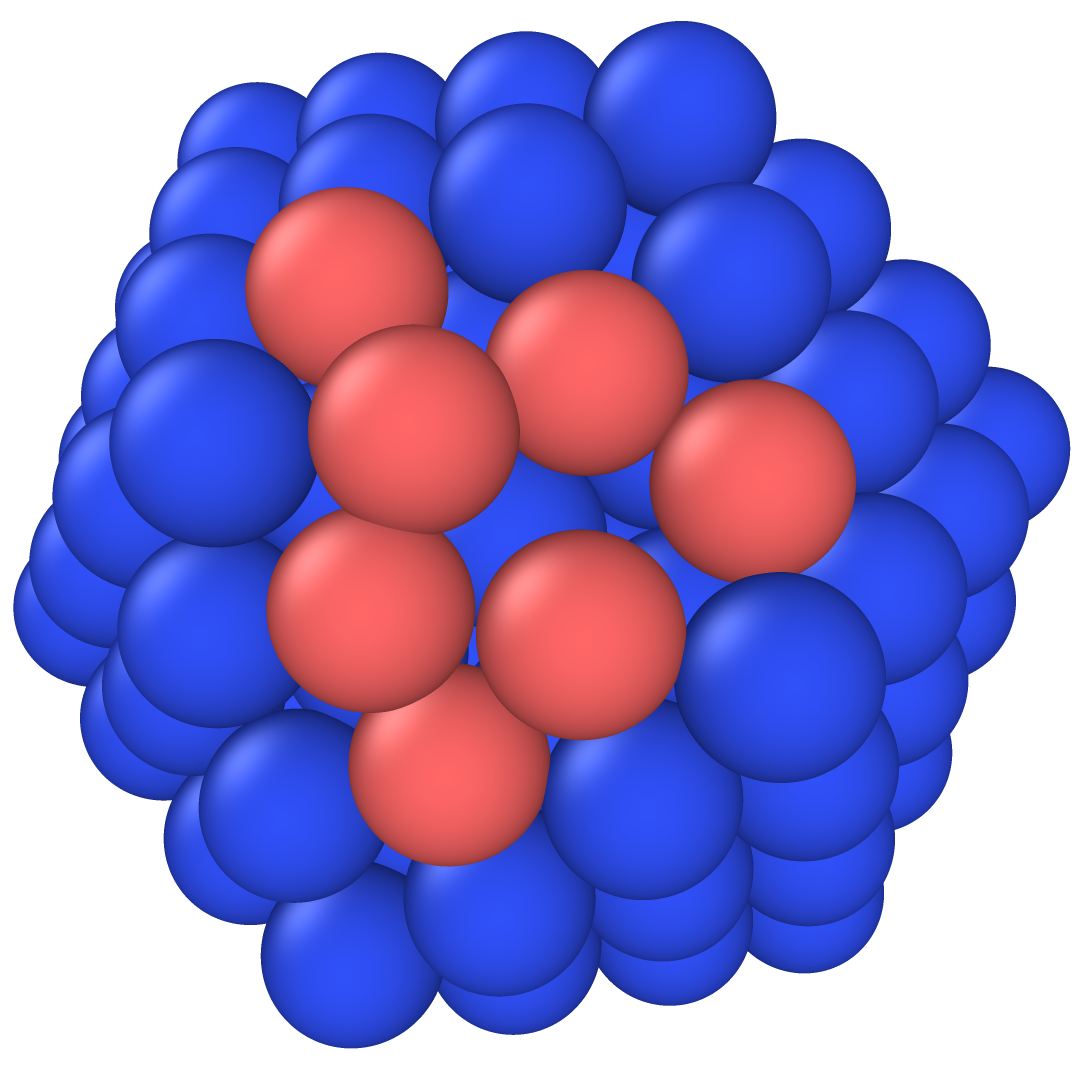} &
\includegraphics[width=0.25\linewidth]{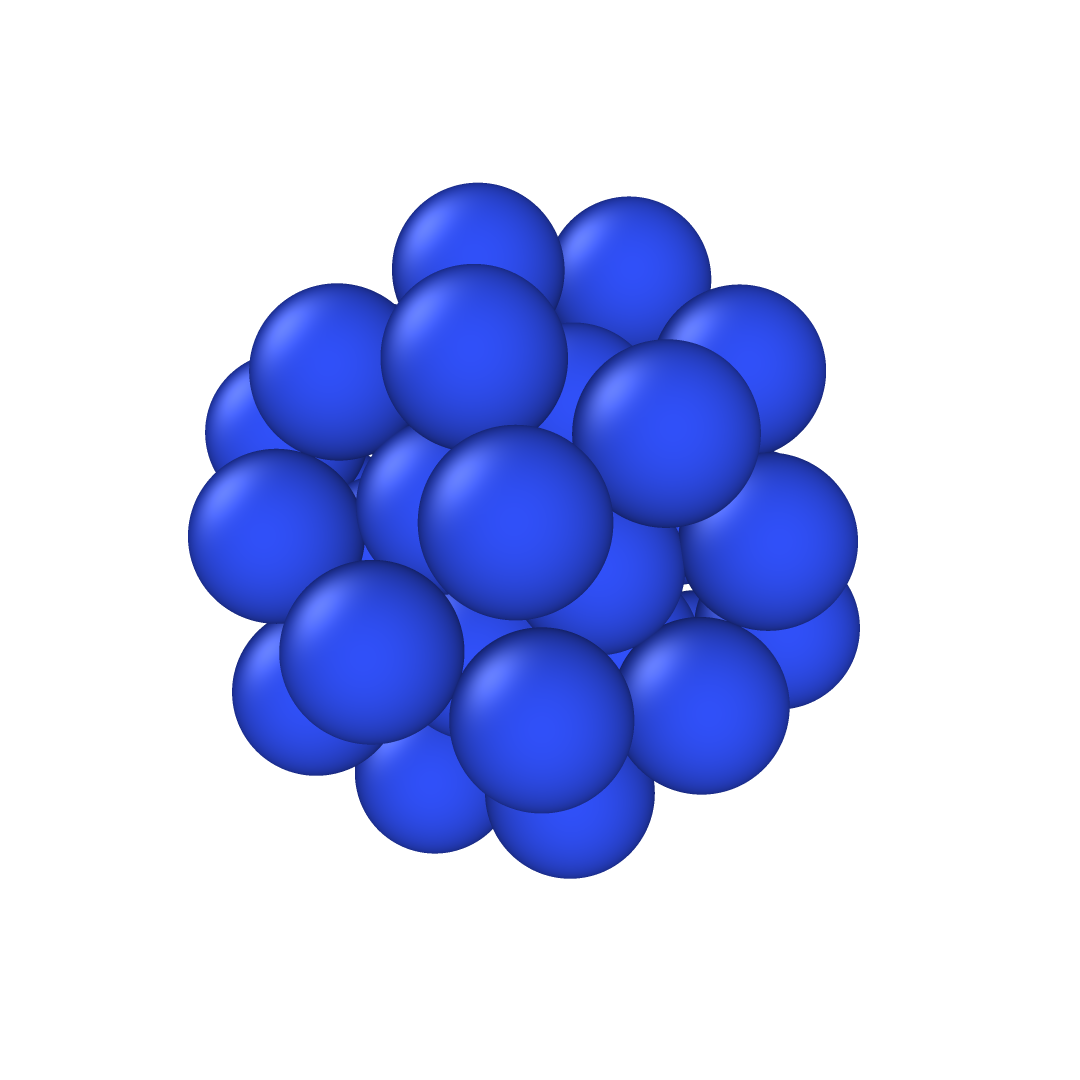} &
\includegraphics[width=0.25\linewidth]{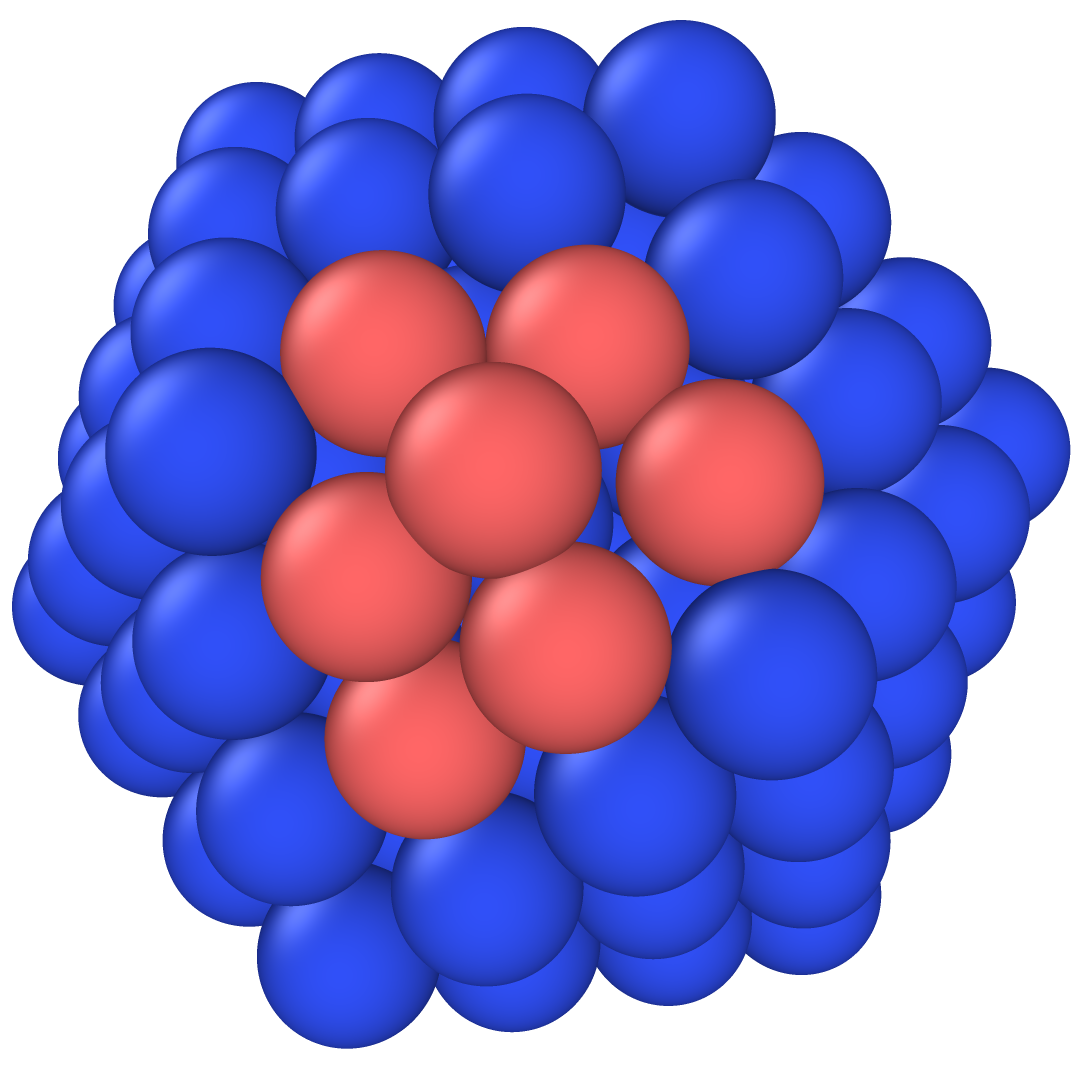} \\[2mm]
\includegraphics[width=0.25\linewidth]{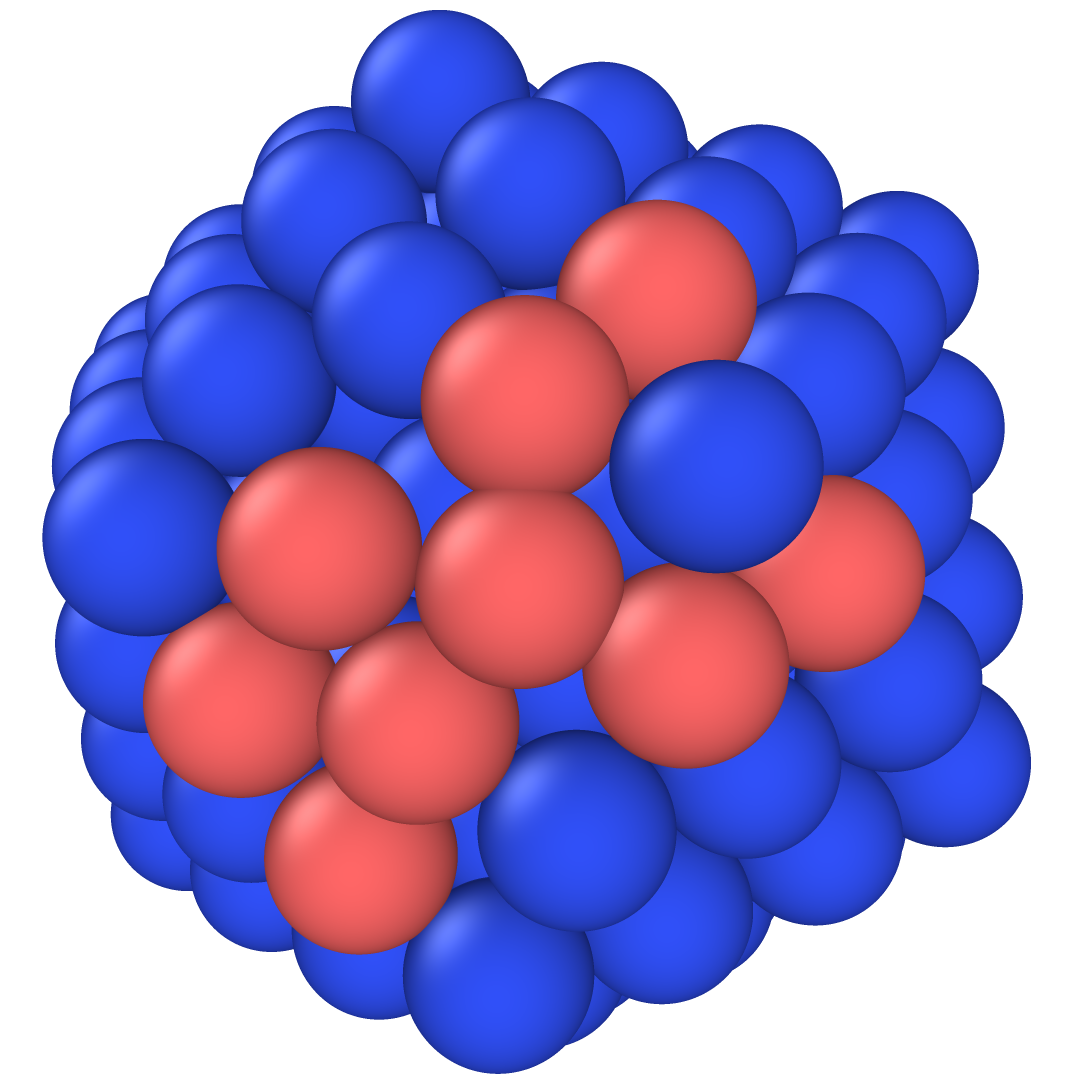} &
\includegraphics[width=0.25\linewidth]{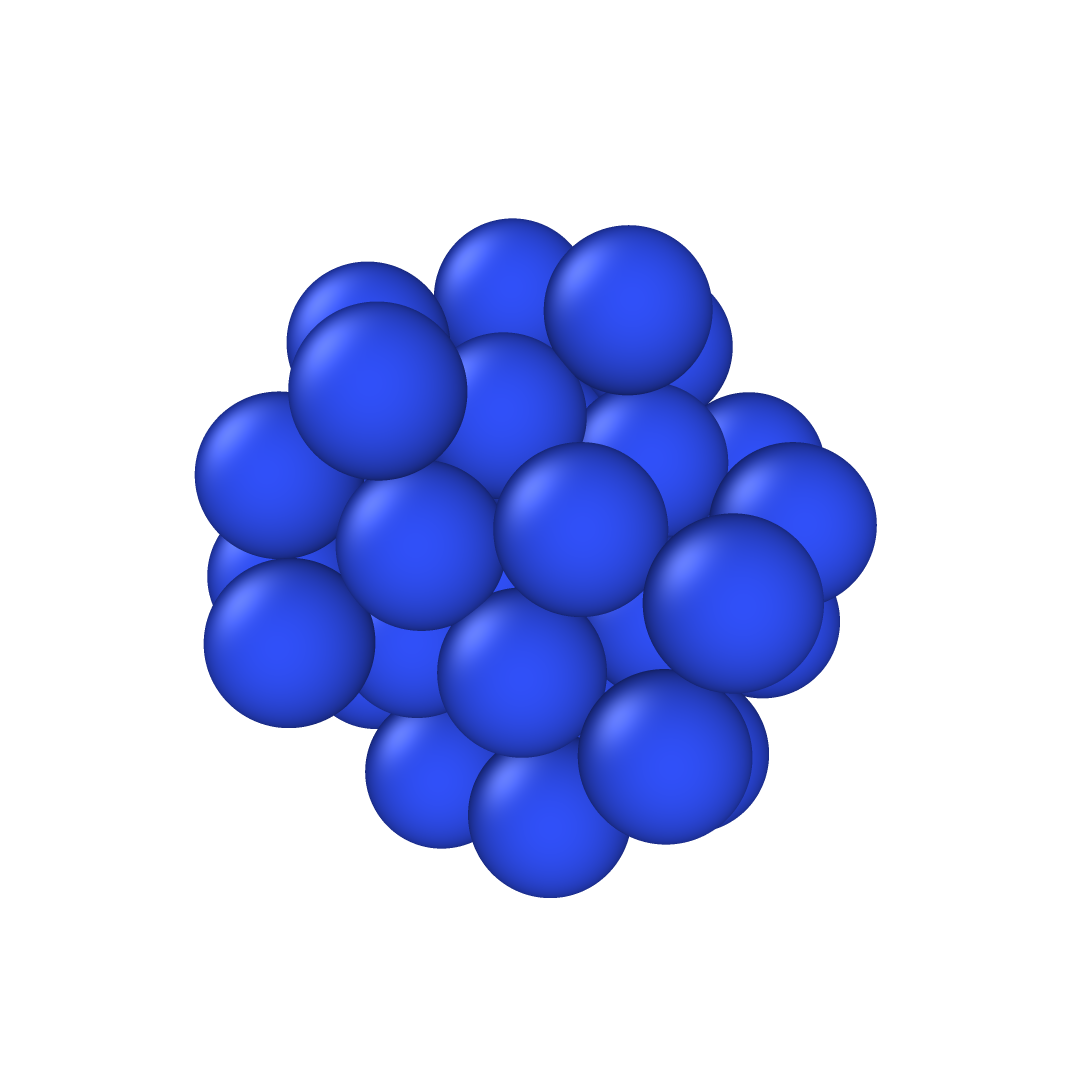} &
\includegraphics[width=0.25\linewidth]{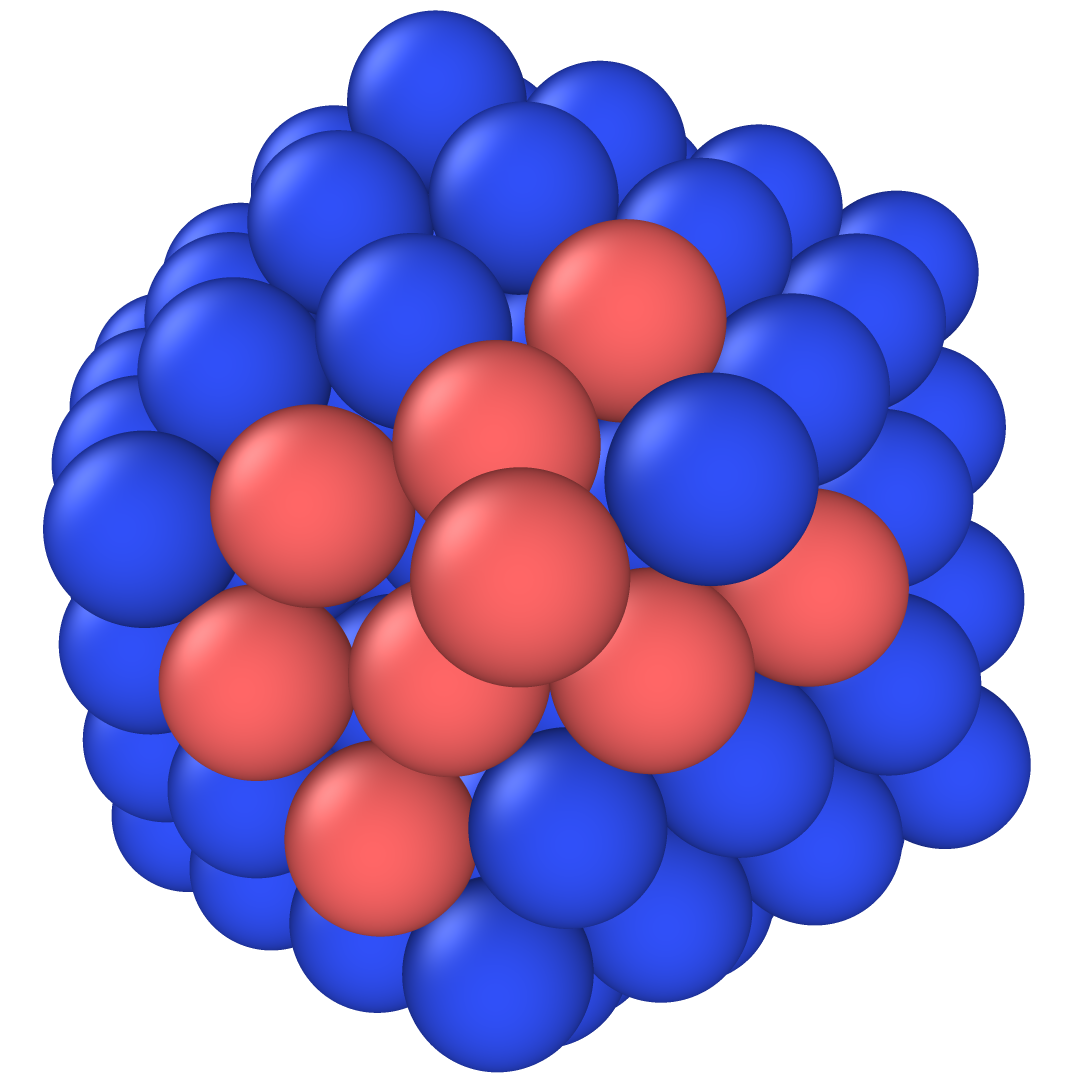}
\end{tabular}
\caption{Five examples of output from the partial-position hinted model. (Left) the initial state provided to the model. (Middle) the partial-position hint given to the transformer. (Right) The correctly predicted final state. ``Hint Sizes'' ($s_h$) from top to bottom are: 7\%, 11\%, 53\%, 15\%, and 13\%.}
\vspace{-1em}
\label{fig:ex_hinted_transitions}
\end{figure}

\begin{figure}
    \centering
    \includegraphics[width=.89\columnwidth]{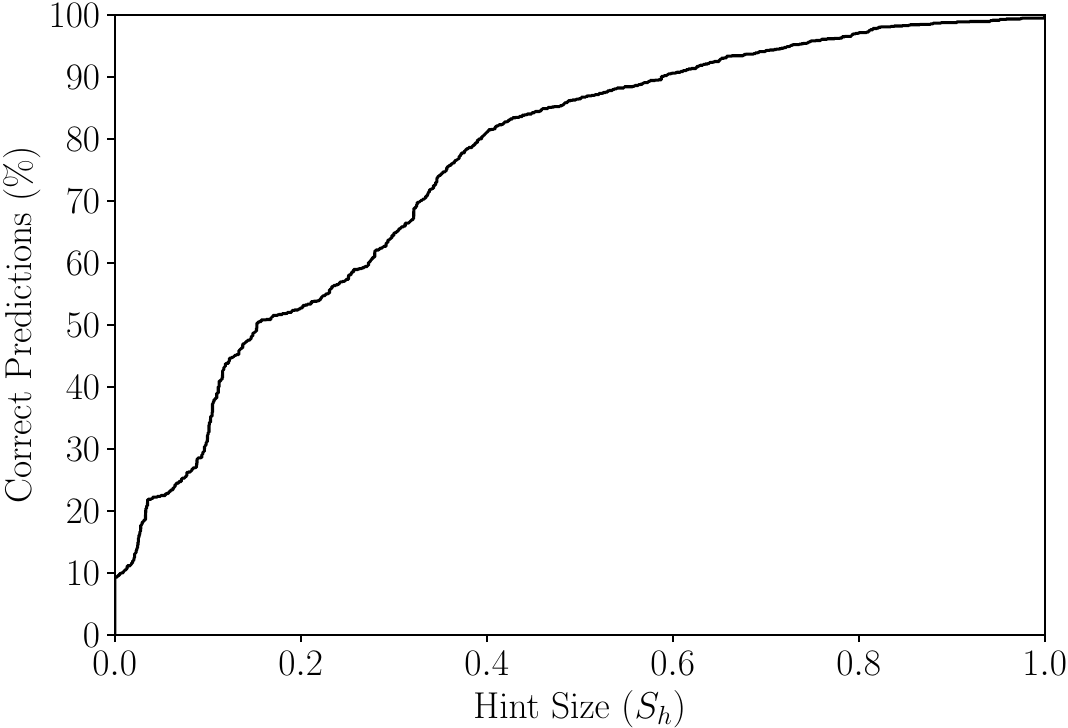}
    \includegraphics[width=.89\columnwidth]{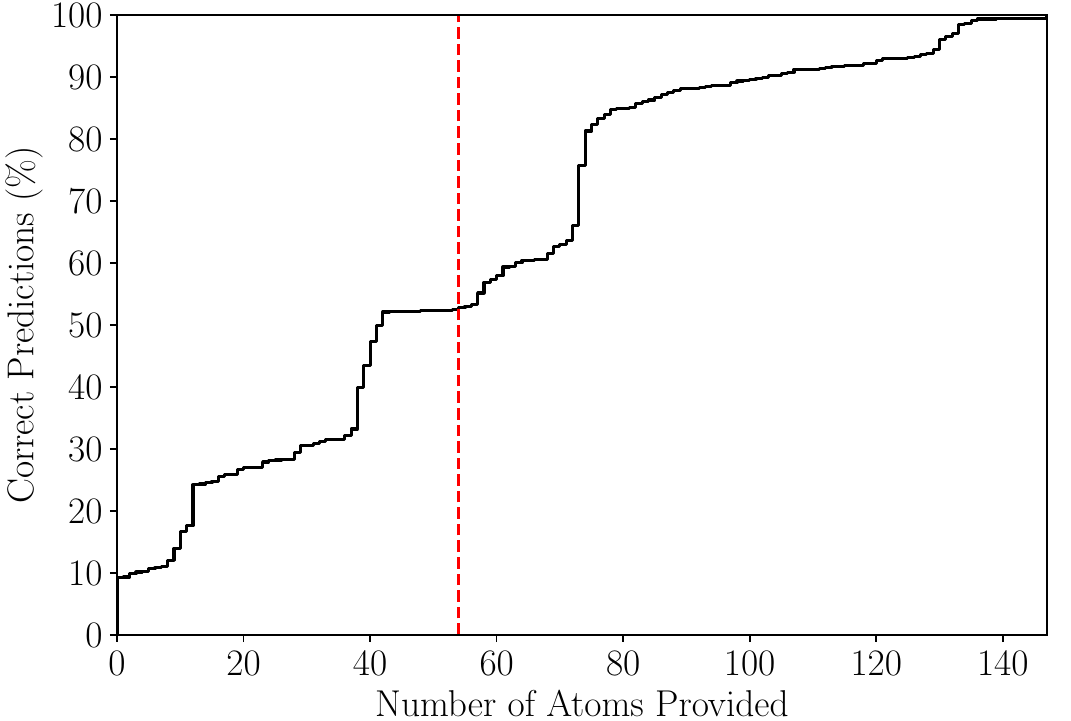}
    \caption{The amount of additional information required for our partial-position hinted model to correctly predict transitions known from our dataset. Top: The percentage of correct predictions depending on the ``hint size'' ($S_h$). Bottom: The percentage of correct predictions based on the number of provided final positions. The dotted line separates the atoms in the surface layer (to the right of the dotted line), where transitions are most likely to occur, from the inside of the cluster (on the left of the dotted line).} 
    \label{fig:min_hint_rdfs}
\end{figure}

We provide some specific examples of partial-position-hinted predictions in Fig.~\ref{fig:ex_hinted_transitions}, where our model is given an initial configuration and is then conditioned with some known final-state positions to suggest a known transition. As shown in Fig.~\ref{fig:ex_hinted_transitions}, this hint is provided from the center of mass and then outwards, meaning that we hint our model with many inner atoms which typically do not transition, before potentially hinting with atoms on the surface of the cluster. We tested 1500 transitions chosen from our validation dataset and found that, on average, we need a ``hint size'' ($s_h$) of 0.25 to correctly predict a known transition. However, we can still predict many transitions with much smaller hints. The proportion of the tested sample of the validation dataset (1500 transitions) that can be predicted for each hint size is shown in Fig.~\ref{fig:min_hint_rdfs}. Note that for a hint size of zero there are about 10\% of the transitions where our model predicts the same final state known from simulation; however, it is difficult to meaningfully interpret this behavior, as there is no particular reason to expect our model to predict exactly that final state we happen to know from simulation and not another, valid state. See below (in particular Sect.~\ref{subsec:res_unhinted}) for more discussions on this topic.


\begin{figure*}
    \centering
    \begin{tabular}{cl}
        \rotatebox{90}{\footnotesize \quad$\Delta$E (eV)} &
        \begin{minipage}{.9\linewidth}
            \includegraphics[width=0.9872\linewidth]{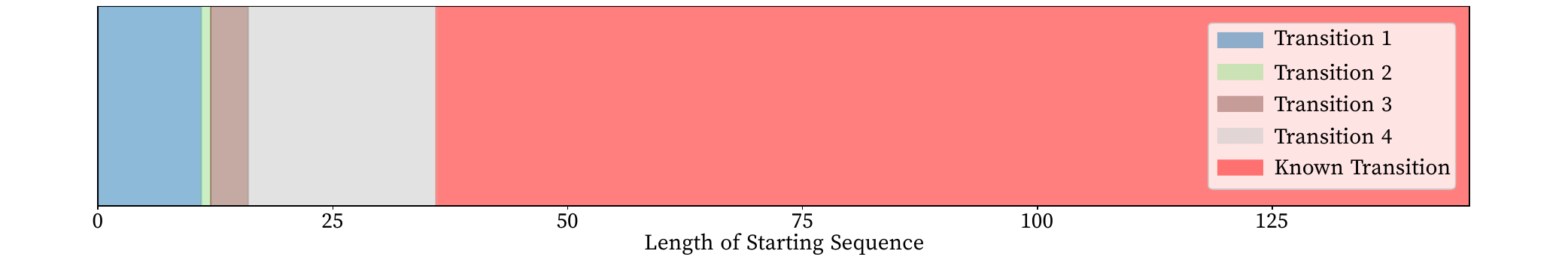}
            \includegraphics[width=0.48\linewidth]{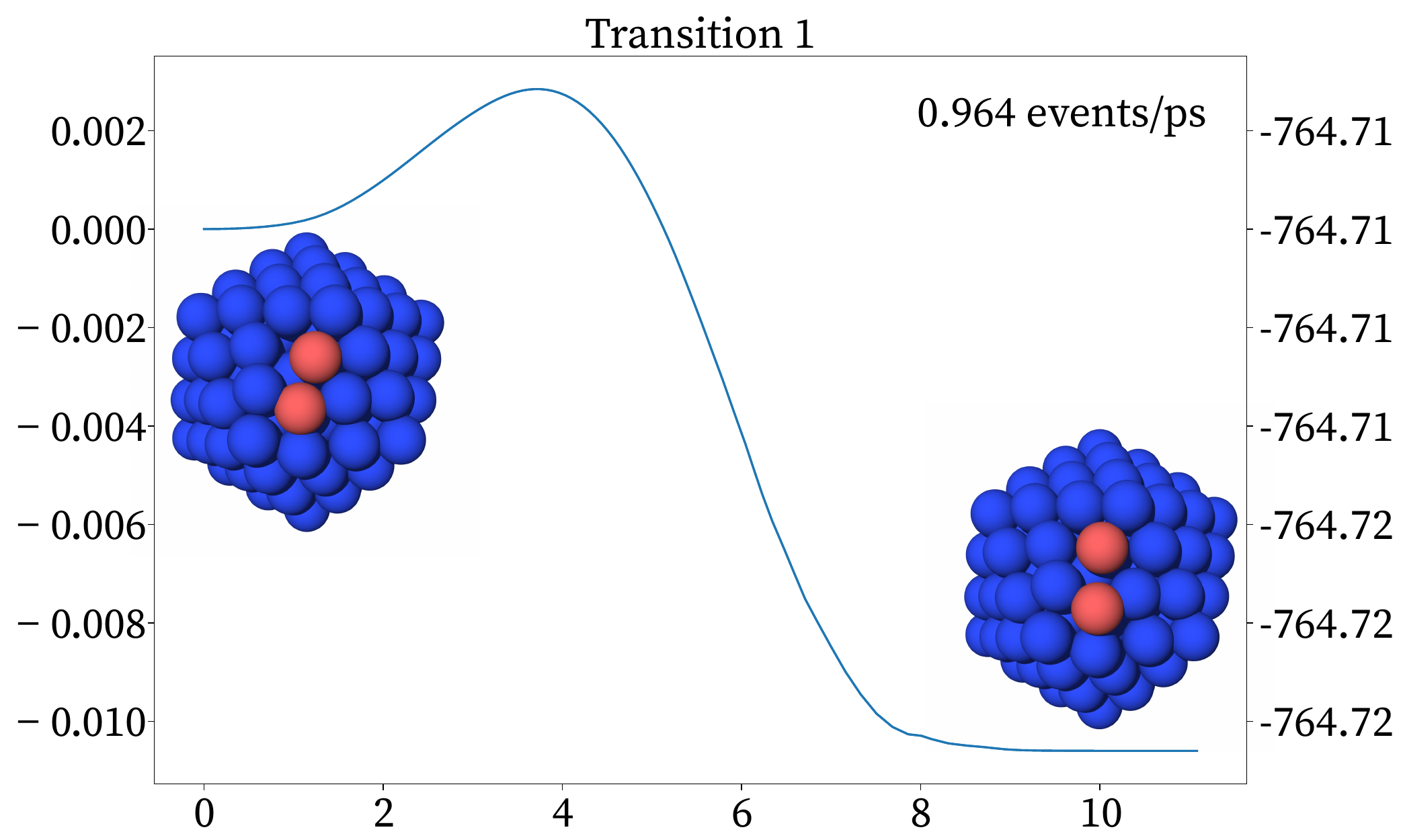}
            \includegraphics[width=0.48\linewidth]{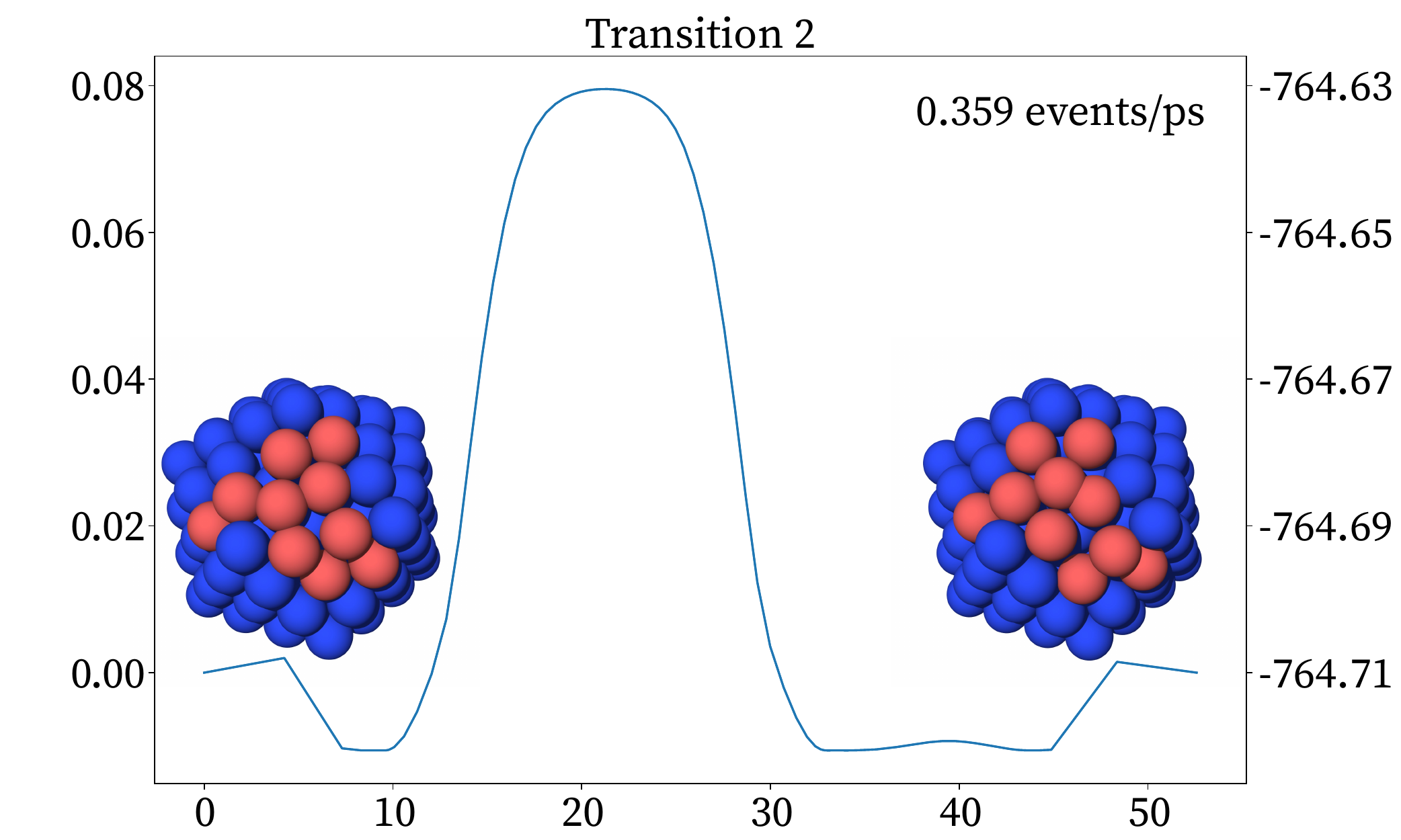}\\[1mm]
            \includegraphics[width=0.48\linewidth]{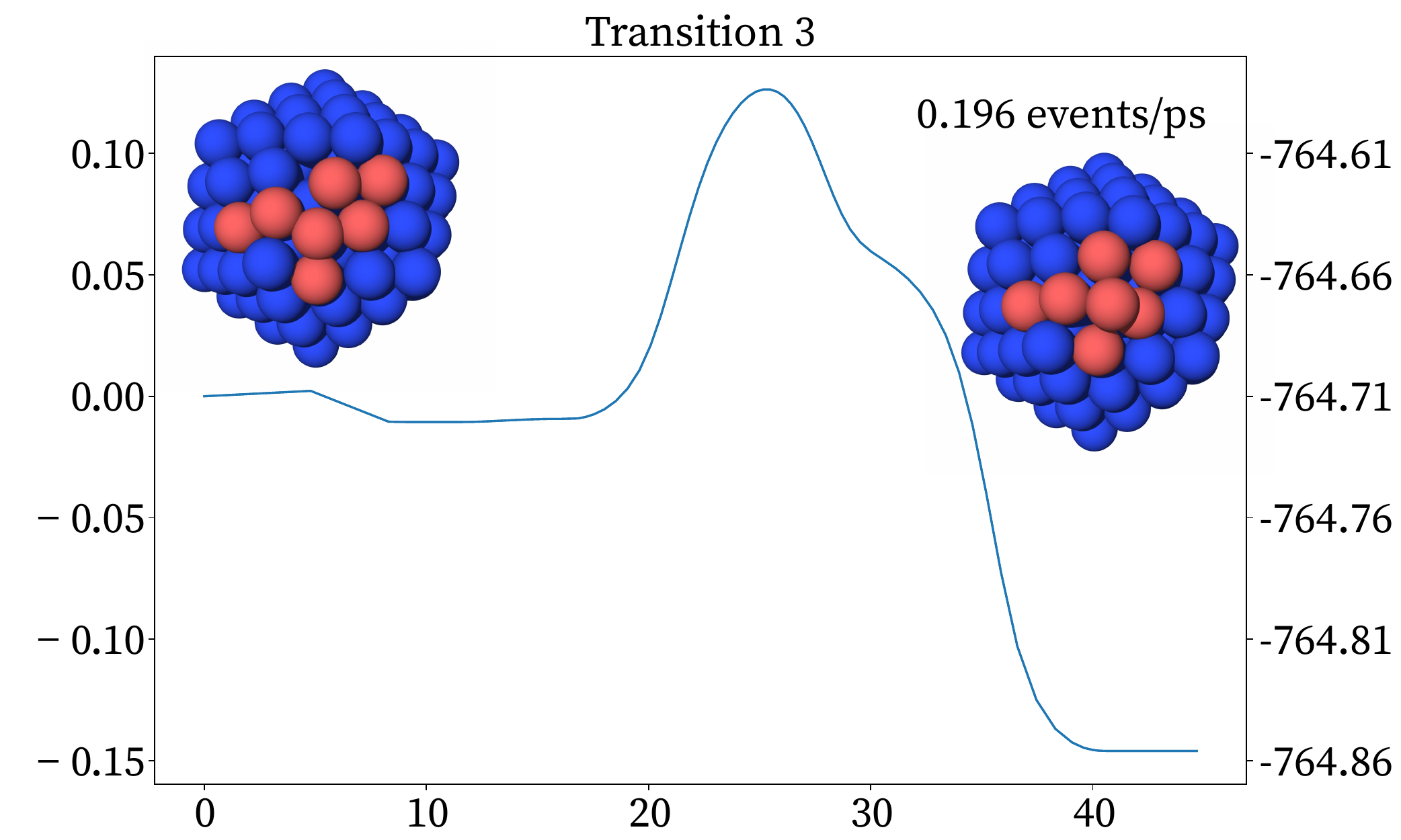}
            \includegraphics[width=0.48\linewidth]{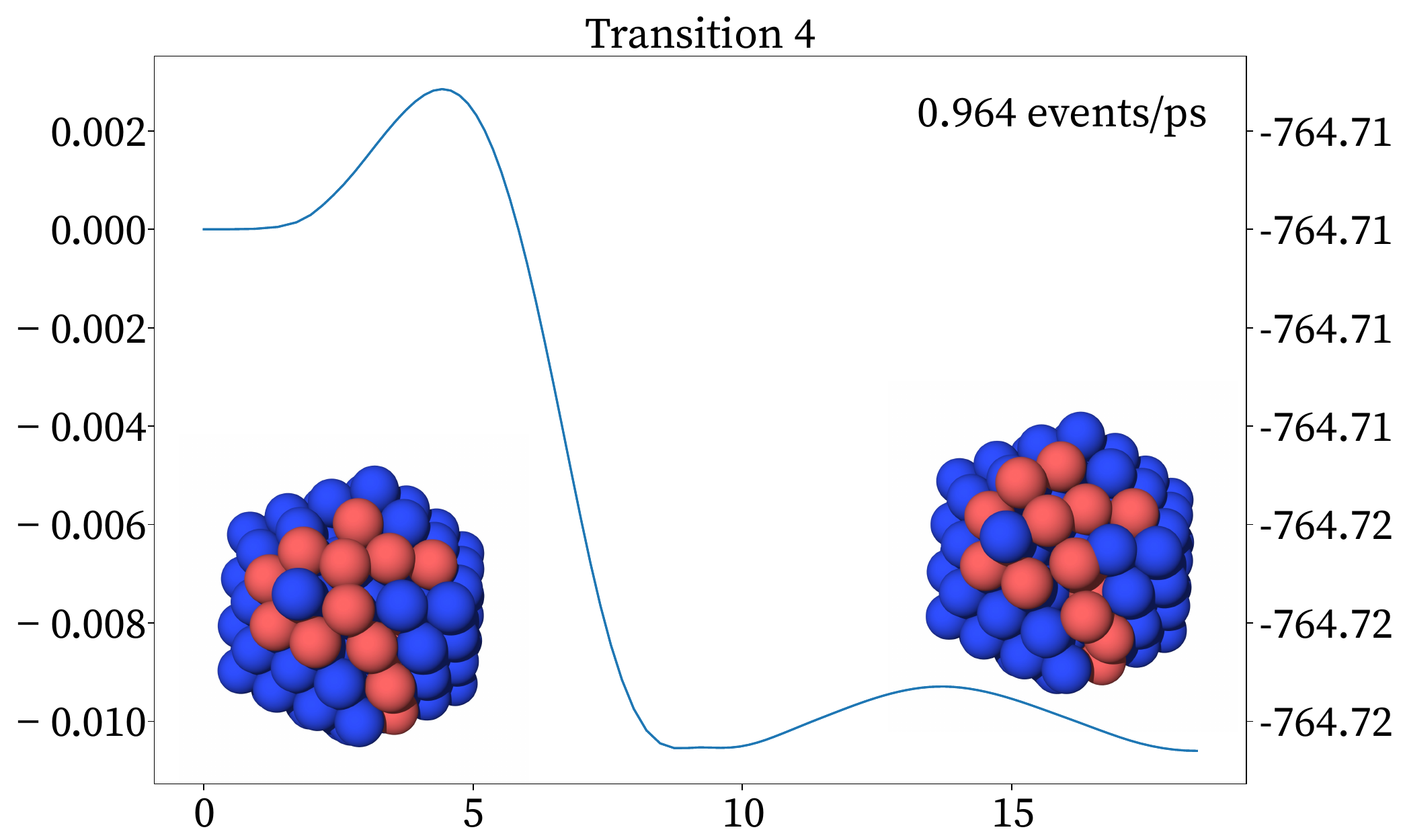}\\[1mm]
            {
                \centering
                \includegraphics[width=0.48\linewidth]{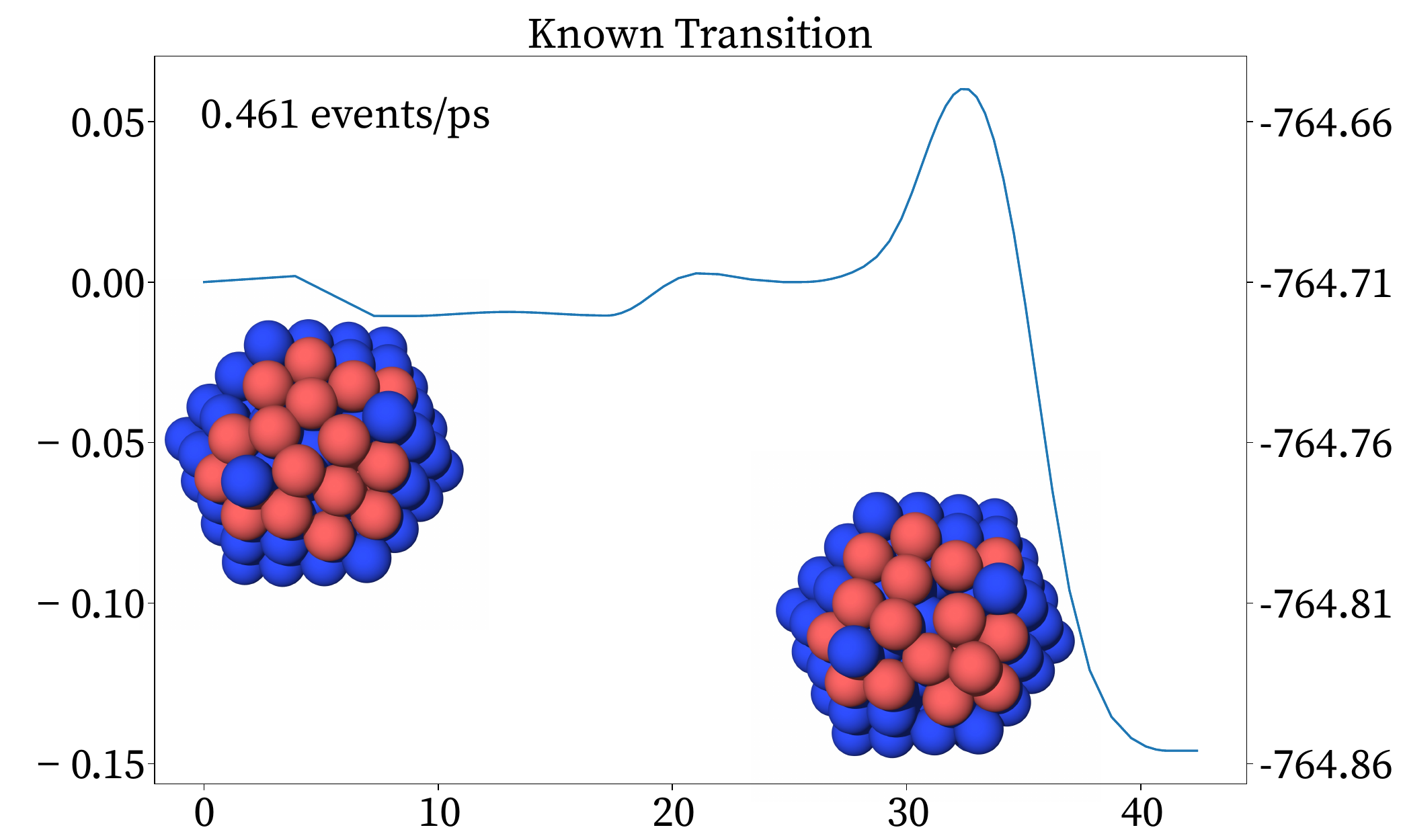}\\
                \footnotesize{Hyperdistance (\AA)}
            }
        \end{minipage} 
        \rotatebox{90}{\footnotesize Energy (eV)} 
    \end{tabular}
    \caption{Transitions predicted by our partial-position hinted model for a single initial state as our hint increases in length. Top: The distinct transition predicted for each possible length of hint provided to our model. Below: Visualizations of these transitions. The initial state (left) and final state (right) are shown alongside a reconstructed minium energy pathway (MEP), calculated using a nudged elastic band (NEB) test. The approximate transition rates are provided in the top right of each plot. Red atoms are those that rearrange during the transition (note that the coloring of the initial state varies between transitions). Hyperdistance gives the cumulative displacement for each frame in the transition pathway.}\vspace{-.5em}
    \label{fig:all-transitions-fig}
\end{figure*}

Although our model typically requires a meaningful hint to generate a specific, known transition, the transitions predicted ``before'' this, as the hint is increased in length, are often dynamically relevant, and can vary multiple times before the final state is ultimately predicted. This behavior is shown in Fig.~\ref{fig:all-transitions-fig}, which gives the various transitions predicted as the length of a hint provided to our model for a particular initial state increases. All of these transitions appear to be physical and dynamically relevant, as seen in their reconstructed transition pathways.

\subsection{Individual Displacement Hinted Model}
\label{subsec:res_displacement_hint}

We also trained a transformer to predict atomistic transitions when provided with an input giving the distance of each individual per-atom displacement between an initial and final state. This additional input is given as an appended column alongside the position of each atom in each initial and final state as follows,
\begin{equation}
     \mathbf{X}^\prime_i = \left(x_i, y_i, z_i,  \lvert \vec{d_i} \rvert,  p_{\{i,1\}}, \ldots, p_{\{i,n\}}\right)
\end{equation}
    where $d_i$ gives the individual per-atom displacement of each atom between the initial and final state and takes the place of one element of $\vec{p}$, which then has a length of \hbox{$n = 156$}. Unlike our model in Sect.~\ref{subsec:res_position_hint} which was given partial-position hints in the final state, this version is trained to expect the magnitude of each individual movement during training. As such, the size of the hint cannot be varied at prediction time. We also note that there is no regularization or implicit bias given to the model that would enforce or suggest these provided magnitudes in the output state.

The model, when hinted like this with the magnitude of each per-atom displacement, very frequently produces the known transition it was conditioned on. This is shown in Table~\ref{tab:individual-movement-hinted-results}, where the model correctly predicts the known transition across approximately 96\% of our validation dataset. To reproduce final states with approximately this same degree of accuracy using the partial-position hinted model, one must provide the model with the first 130 atoms of the known final state, or a \smash{``hint size'' ($s_h$) of 0.75 (see Fig.~\ref{fig:min_hint_rdfs}).}
\begin{table}
        \centering
        \caption{Model prediction outcomes for the individual displacement hinted model. ``Correct'' refers to states with an identical connectivity graph compared to the known final state. ``Incorrect'' refers to the case when the model predicts a final state different from the expected one in that regard. ``Fail'' means the predicted final state is equal to the initial state (that is, no transition is predicted).}\vspace{0.5em}
        \newcolumntype{C}[1]{>{\centering\arraybackslash}p{#1}}
        \begin{tabular}{|C{2cm} C{2cm} C{2cm}|}
        \hline
        Correct & Incorrect & Fail\\ \hline
        &&\\[-.25cm]
        96\,\% & 2\,\% & 2\,\% \\ \hline
        \end{tabular}
        \label{tab:individual-movement-hinted-results}
\end{table}

\subsection{Autonomous Predictions}
\label{subsec:res_unhinted}

Previously, we have shown that our model can reproduce transitions known from simulation with small --- and even without --- suggestion, and can predict reasonable and distinct transitions at varying hint sizes $>0$. We will now show how our model can in fact make autonomous predictions of multiple final states from an initial state alone, that is, without any hints or knowledge of final states.

\begin{figure}
    \centering
    \begin{tabular}{@{}c@{\hspace{2pt}}l@{\hspace{2pt}}c@{}}
    \rotatebox{90}{\footnotesize $\Delta$E (eV)} &
    \begin{minipage}{0.88\linewidth}
        \includegraphics[width=\linewidth]{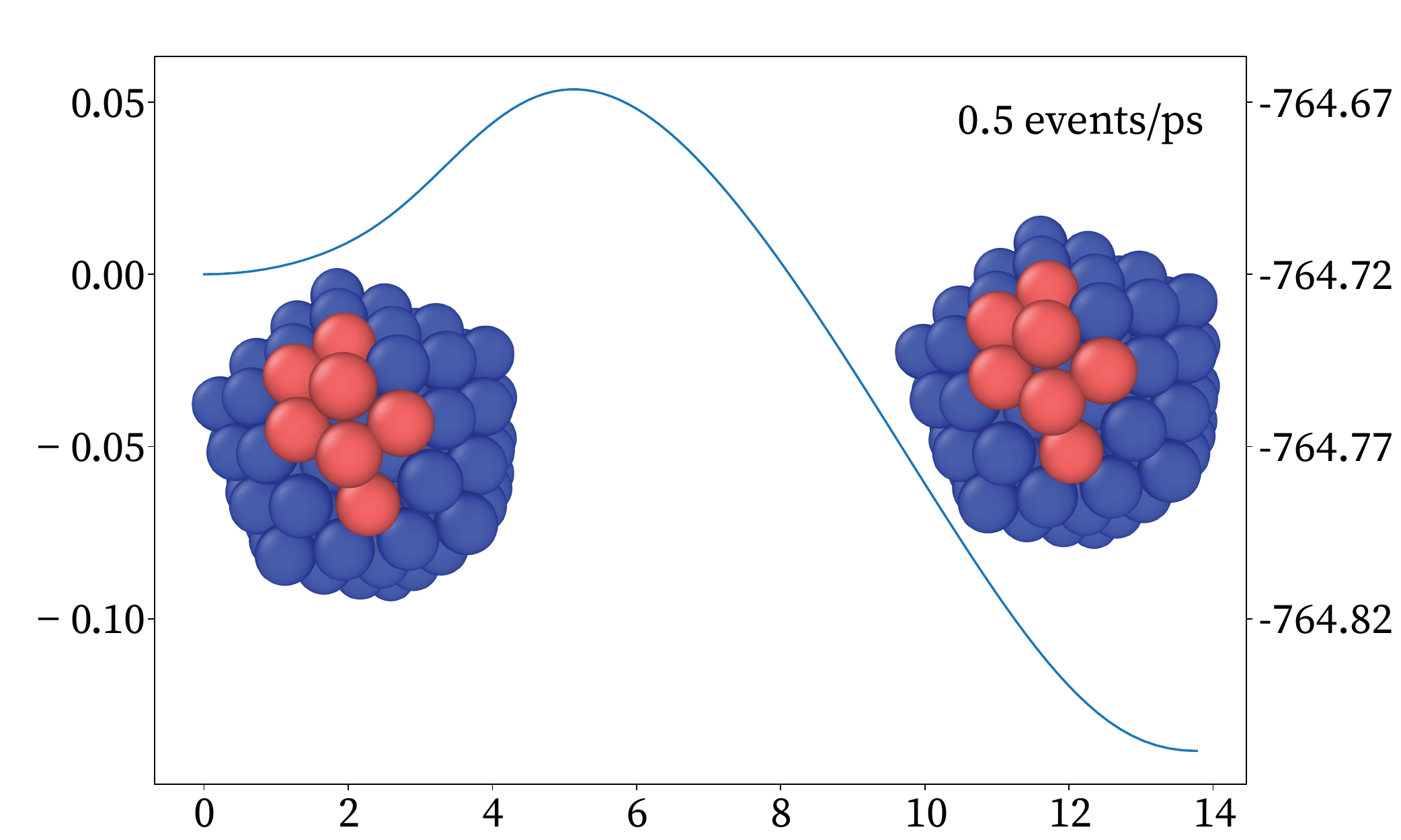}\\
        \includegraphics[width=\linewidth]{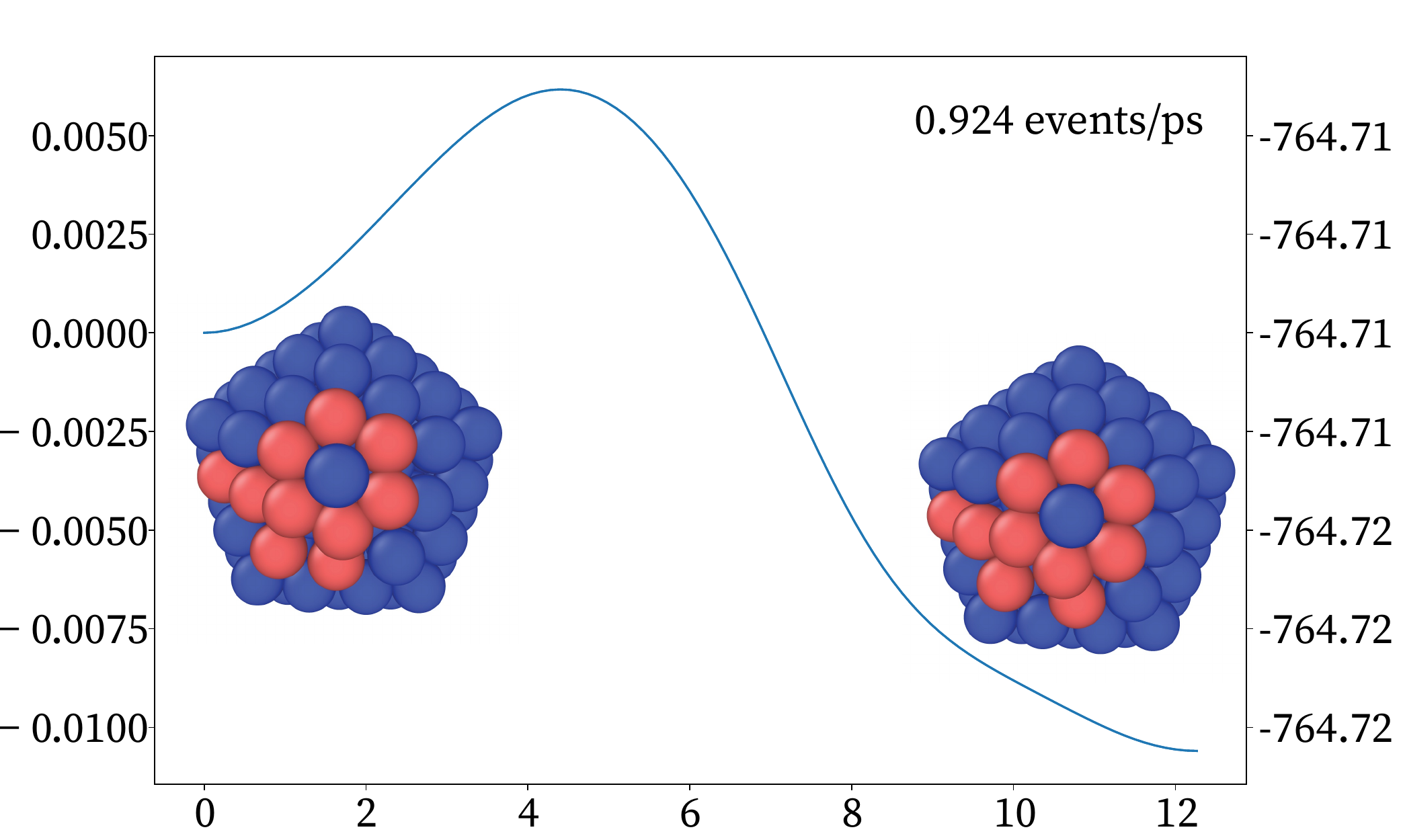}\\
        \includegraphics[width=\linewidth]{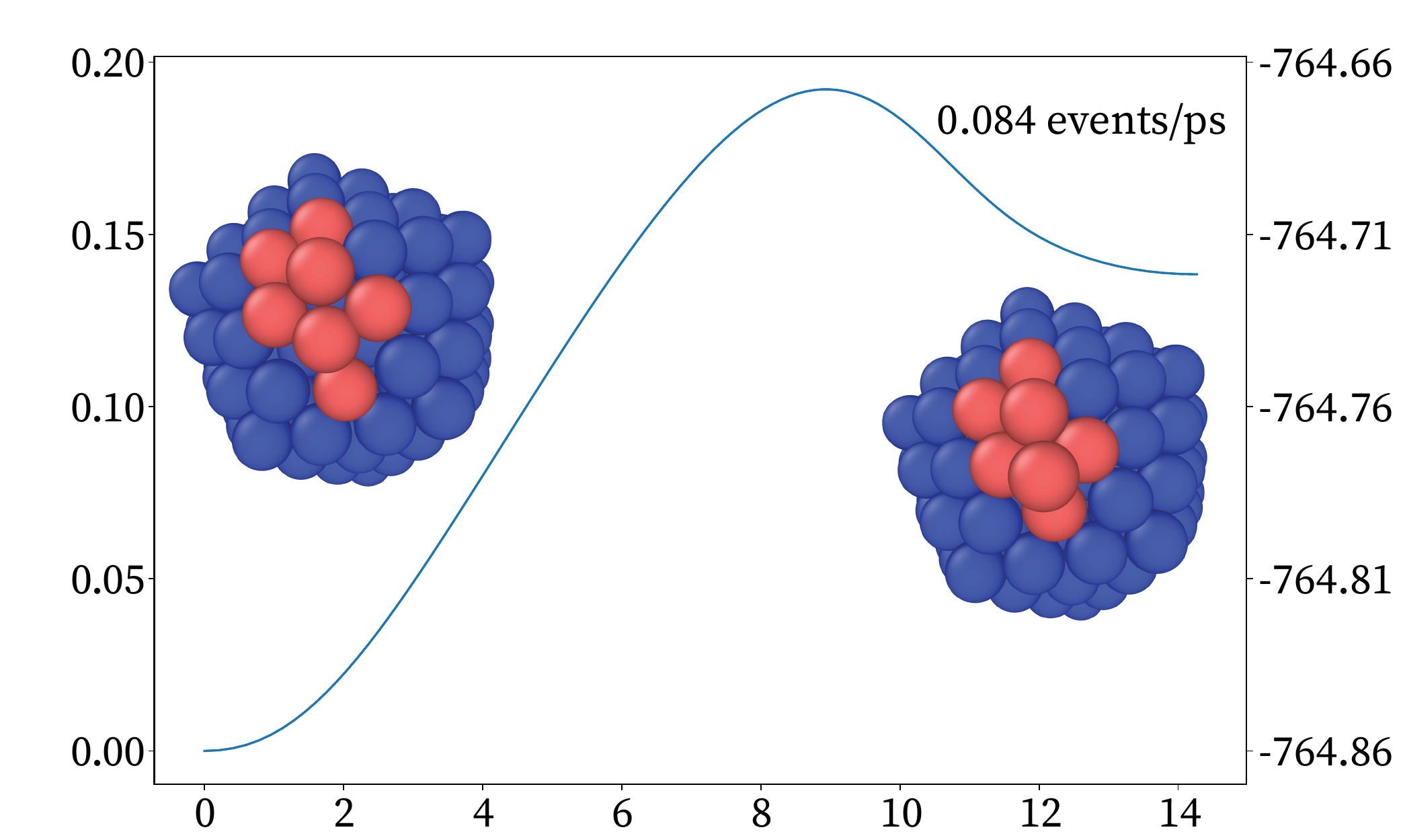}\\
        \centering \footnotesize Hyperdistance (\AA)
    \end{minipage} &
    \rotatebox{90}{\footnotesize Energy (eV)}
    \end{tabular}
    \caption{Selected examples of predictions made by our model without any additional hints. Left: Visualizations of the initial state. Right: Visualizations of the final sate. Red atoms emphasize those that rearrange during the transition. Plots contain the Minium Energy Pathway (MEP) for our prediction, calculated using a Nudged Elastic Band (NEB)~test.}
    \label{fig:unhinted-nebs}
\end{figure}

We first show that our ``unhinted'' model is capable of predicting accurate transitions for many initial states in our dataset. We show examples of NEB calculations for unhinted outputs (that is, predictions using our ``partial-position hinted'' architecture though without any additional suggestion) in Fig.~\ref{fig:unhinted-nebs}, all of which give physical and dynamically relevant transition pathways

Now, motivated by the ability to predict varying final states as different prior information is provided, shown before in Fig.~\ref{fig:all-transitions-fig}, we will discuss a specific method of autonomously generating multiple transitions from a single input state, including previously unknown ones. We emphasize that we do not provide any knowledge about final states anymore, as we have done above. Instead, we provide our model with small, random perturbations of the initial state, as it would be the case in a realistic scenario where final states are unknown.
To do so, we first perturb a small set of ``inner'' atoms in the initial state configuration by adding random noise to their positions. For the experiments described here, we chose the 20 atoms closest to the center of mass of the cluster. Then, we use our transformer model to predict a final state, using this perturbed section of the initial state as a prior for the final state (in place of a known partial-position hint). We repeat this procedure multiple times and then identify identical transitions within the set of ``raw'' predictions.

\begin{figure}
    \centering
    \includegraphics[width=\columnwidth]{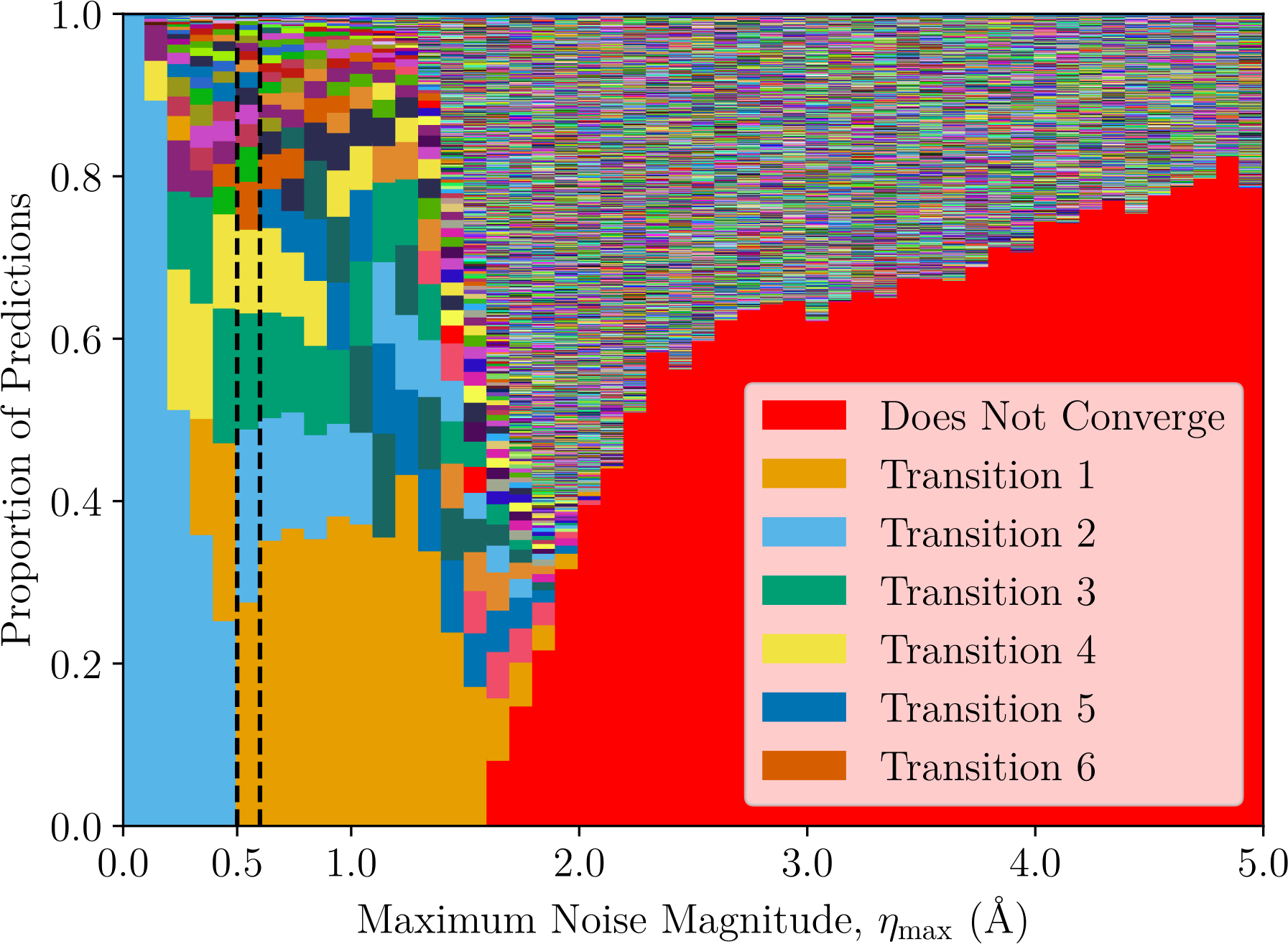}
    \caption{The transitions predicted with our perturbation method as the magnitude of the noise added to the core of the initial state is increased. Solid colored bars represent the number of times a specific transition was predicted, the red area indicates the proportion of final states that failed to converge to a metastable state. Visualizations of the most common predicted final states using noise with $\eta_{\mathrm{max}}=0.5$\,{\AA} (highlighted) are shown in Fig~\ref{fig:pertubation-distribution}.}
    \label{fig:transitions-with-noise}
\end{figure}

\begin{figure*}
    \centering
    \begin{tabular}{cl}
        \rotatebox{90}{\footnotesize{$\Delta$E (eV)}} \hspace{-1em} &
        \begin{minipage}{0.95\linewidth}
            \includegraphics[width=.49\textwidth]{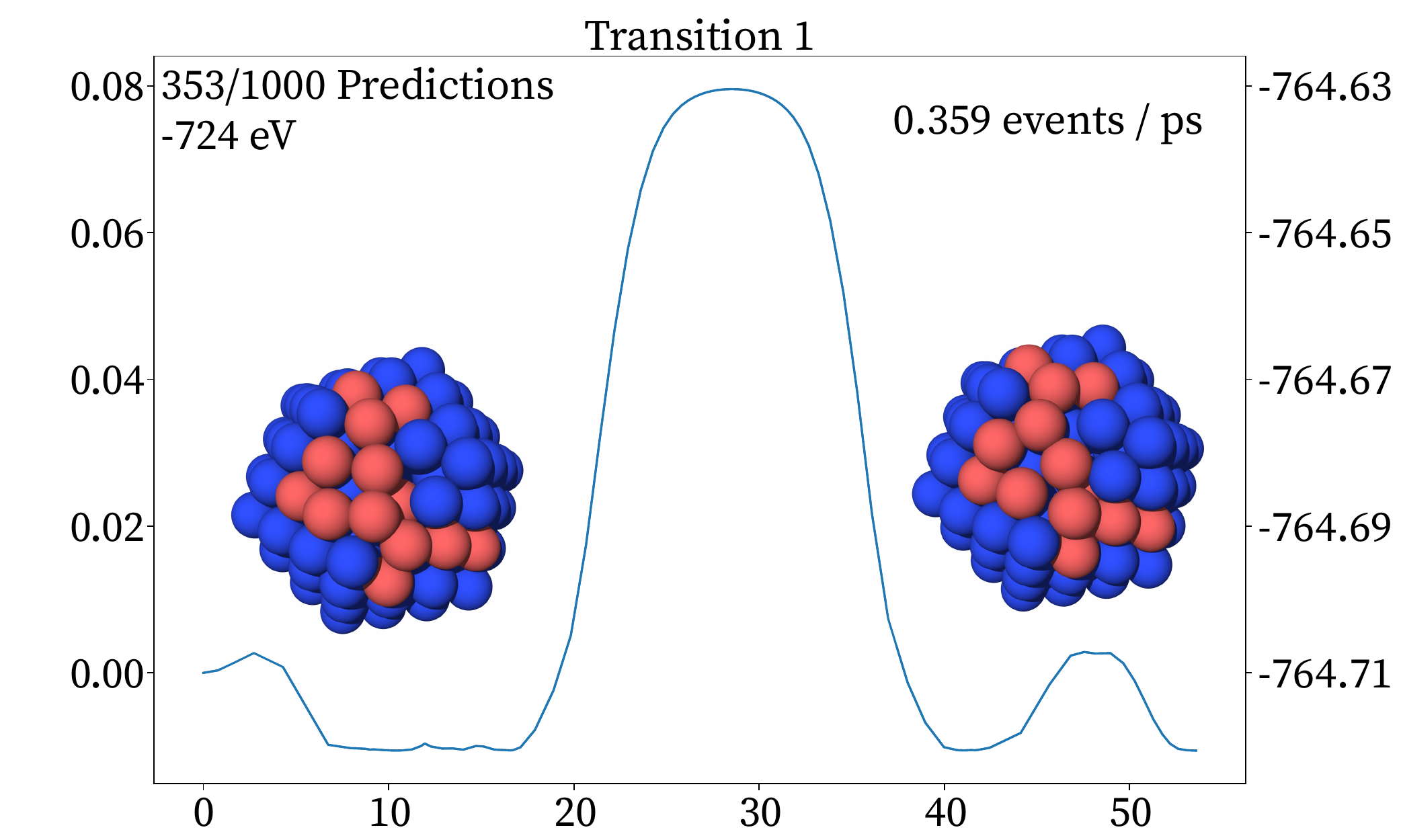}
            \includegraphics[width=.49\textwidth]{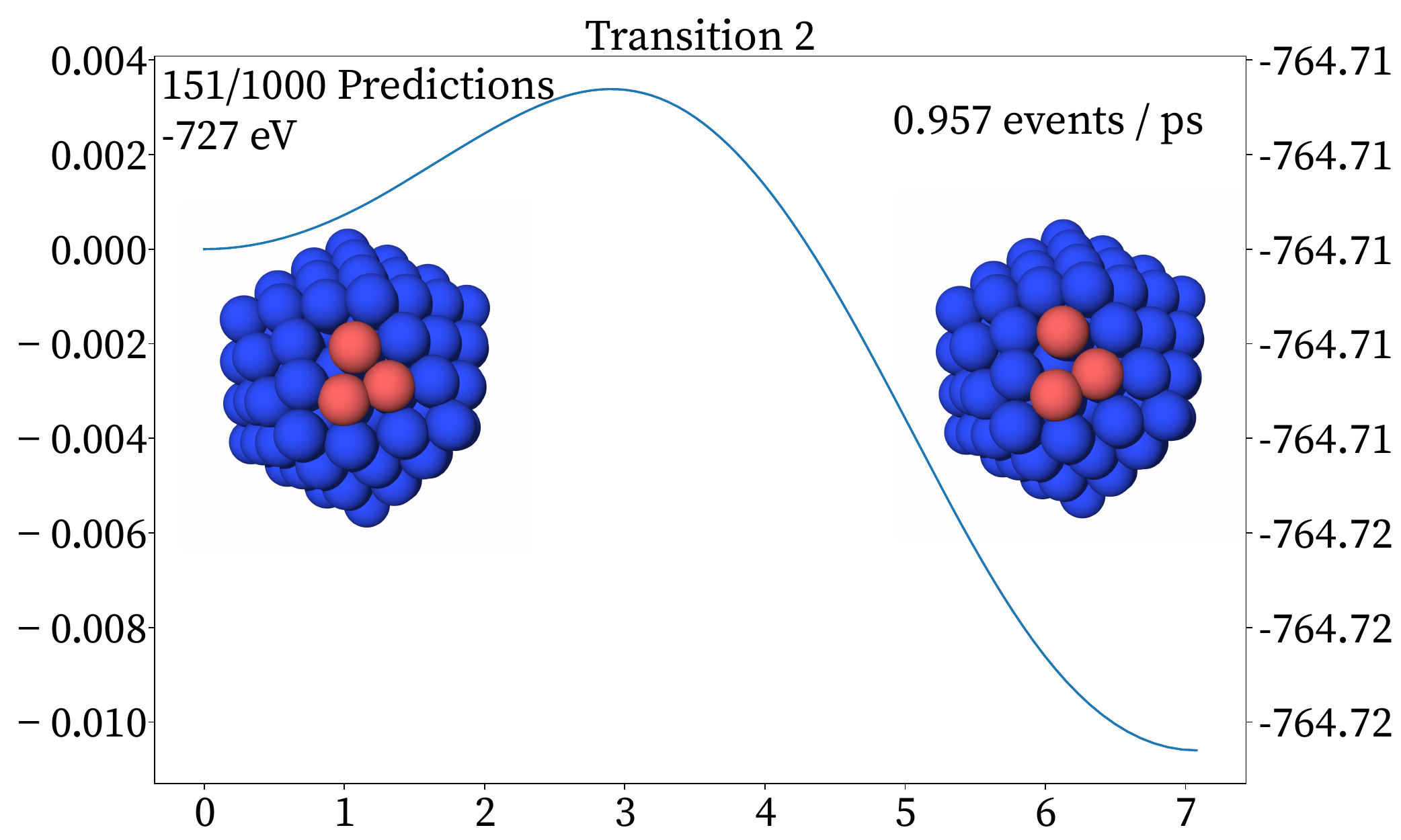}\\
            \includegraphics[width=.49\textwidth]{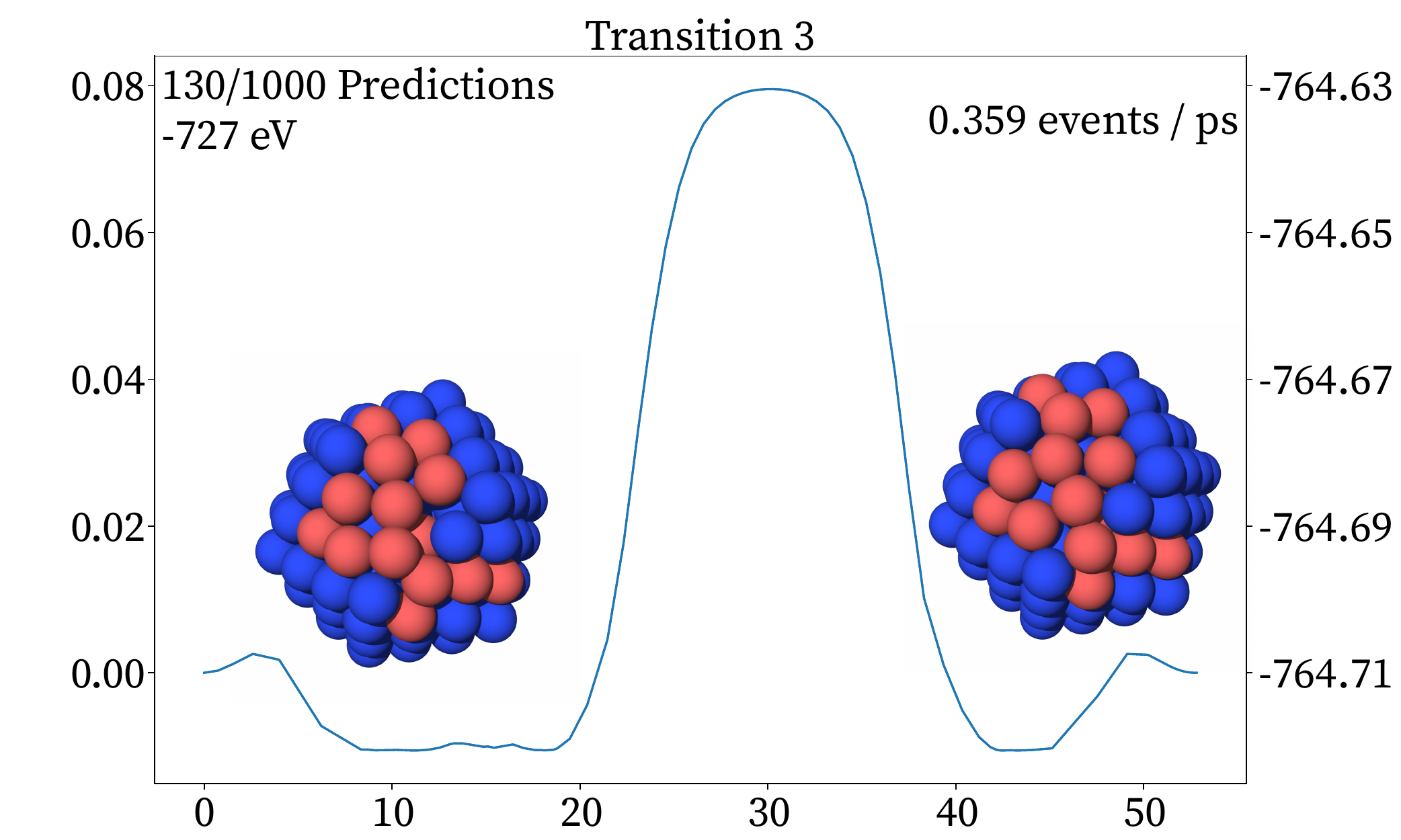}
            \includegraphics[width=.49\textwidth]{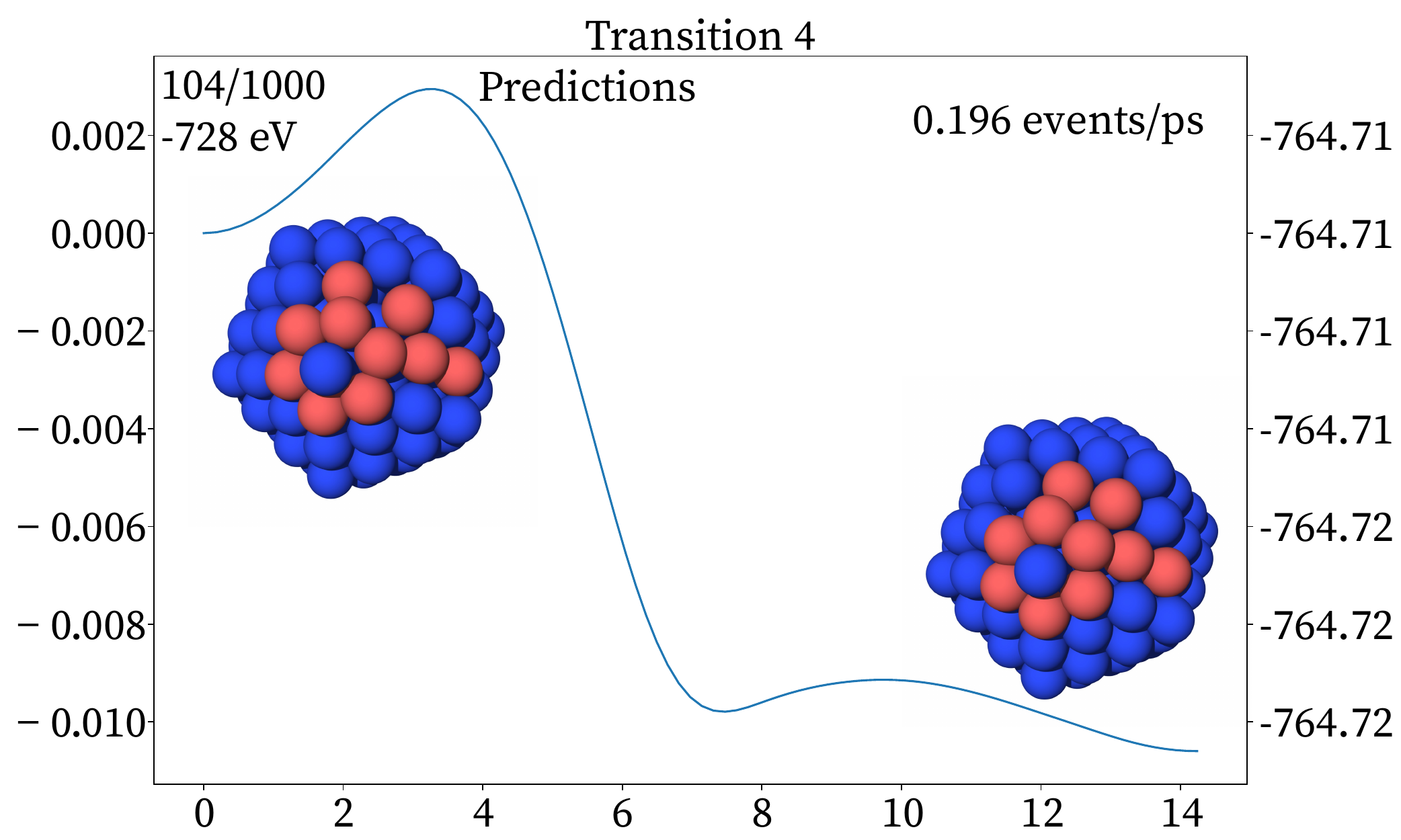}\\
            \centering \footnotesize{Hyperdistance (\AA)}
        \end{minipage} 
        \rotatebox{90}{\footnotesize{Energy (eV)}}
    \end{tabular}
    \caption{The four most common transitions predicted by perturbing the innermost 20 atoms in the initial state with random noise of $\eta_{\mathrm{max}}=0.5$\,{\AA} (cf. Fig.~\ref{fig:transitions-with-noise}). In the top left corner of each plot we show the number of times the transition was observed and the average energy of the predicted final state before optimization. In the top right corner of each plot the approximate transition rate is given. None of these transitions were previously known from our simulations. Note that transitions 1 and 3 are very similar, but not identical.}
    \label{fig:pertubation-distribution}
\end{figure*}

The magnitude of the noise, $\eta$, used to perturb the ``core'' of the initial state significantly affects the predicted final states. One can imagine that without added noise, the model would be forced to make the same deterministic prediction; with large noise the given prior would have little resemblance to any actual physical final state, forcing the model to make mostly unphysical predictions. For priors consisting of the innermost 20 atoms, the quantitative relationship between the magnitude of noise used to create our perturbed prior and the occurrence of certain predicted final states is shown in Fig.\ref{fig:transitions-with-noise}. We will limit further discussion to the final states predicted by adding noise with a magnitude $0.0 < \eta \leq 0.5\,\mathrm{\AA} = \eta_{\mathrm{max}}$, which we believe provides a reasonable balance between stability and the number of predicted final states. One can certainly tune these priors further, both in the number of atoms provided and the magnitude of the noise used to perturb them, to produce other sets of predictions. The final states most frequently predicted with these parameters are shown in Fig.~\ref{fig:pertubation-distribution}; each of which appear to be physical and dynamically relevant.

We emphasize that the purpose of this experiment is not to present a practically competitive tool for enumerating complete sets of transitions from a particular initial state. Rather, by demonstrating that we are able to autonomously predict multiple relevant transitions from a single initial state, with final states that are close to their low-energy configuration already and require little further optimization, we present what we believe is a proof-of-concept for further development of surrogate models to predict these distributions of predicted final~states.

\section{Discussion}

We present a model which has ostensibly learned key properties of transitions, such that it can predict physical transitions and be suggested towards predicting specific transitions known from simulation.  This we believe constitutes a proof-of-concept demonstrating that further development of deep learning methods could produce an effective tool for transition discovery.

Although guaranteeing true statistical correctness from the kind of model we present here is an ambitious goal, which we do not believe to directly show\footnote{The truly rigorous way to measure the statistical correctness is to generate very long explicit dynamical trajectories to estimate the relative probability of each possible pathway. These can then be compared to the predictions of the model to assess whether certain high-probability pathways are missing from the generative predictions, etc. This is technically straightforward in principle, but is likely to be extremely time consuming in practice given that assessing completeness of transition catalogs is a complex and computationally intensive problem~\cite{swinburne2018self}.}, we also find it generally unlikely that our model has an extreme bias toward any specific types of transitions, and rather has a roughly generalizable ability to predict the different kinds of transitions which may occur in our physical system. This is demonstrated in our models highly suggestible nature towards a variety of known final states, alongside the alignment of our predicted out-of-dataset transitions with the physical properties of our transitions known from simulation.

We find that for many transitions in our dataset, our model can be suggested towards predicting the known final states. The amount of information that needs to be provided for these different transitions to be ``accurately'' predicted varies greatly across our dataset, and it is very difficult to quantify how ``obvious'' a transition should be when one is provided with some of the known final state positions. For many known transitions, it is obvious that our model must learn a great deal, as virtually no significant information about the known final state is provided. And, although far less common, some transitions require a hint providing the majority of the movements in the transition to be predicted accurately. 

For transitions which require hints in-between these two extreme modes, it is difficult to rigorously analyze the extent to which our model is meaningfully predicting information about the transitions, or is instead simply completing details which are ``obvious'' when provided with the given hint, though extremely difficult to guess in practice. However, as our model can predict the majority of transitions with less than half of the total atoms in the final state (the majority of which are likely not substantially involved in the transition) and can typically do so with a much smaller part of our ``hint size'' ($s_h$) we believe that our model, although not demonstrating that all of these transitions can be predicted autonomously, has learned key properties about many types of transitions. 

Further, the out-of-dataset transitions predicted by our model all appear to be physical and dynamically relevant. Specifically, the majority of predicted transitions not found in our dataset have reasonable transition pathways, and contain sub-millisecond transition rates similar to those from transitions found during simulation. Our model can often predict these unseen transitions which still obey physical properties when provided with incomplete partial position hints -- which can significantly constrain the possible final states a configuration can transition to. From this, we show that our model is not biased to the point that we are only able to predict final states along one particular prediction pathway -- rather, our model can make accurate predictions even when its prediction branch is constrained.

The model presented here is certainly not a rigorously unbiased, statistically correct predictor of transitions. However, we show that our model can predict physical transitions from an initial state alone, predict additional physical transitions as the models prediction branch is constrained, and reproduce known transitions from minimal information about their final state geometries. As such, we believe to present a non-comprehensive, though generally favorable proof-of-concept for the use of similar machine learning models as surrogates to predict atomistic transitions. 

\section{Summary and Outlook}
This work presents a transformer model that can correctly reproduce known atomistic transitions when provided with minimal additional information about the final state geometry. Furthermore, it can autonomously predict a multitude of previously unknown relevant transitions when the input is slightly disturbed with random noise. These abilities provide a powerful proof of concept for more complex surrogate models which could, for example, assist in saddle search by predicting transitions with random seeding, or for probabilistic models which can sample the transition space directly. Such surrogates bear the potential to ultimately replace conventional large-scale numerical methods, allowing materials simulations to run with a vastly reduced computational cost and increased scope.

Even though we show that a deep-learning model (specifically a transformer) can learn key properties of transitions, the possibility of a bias in the ensemble of transition pathways predicted by our model remains. That is, we cannot exclude the possibility that our model misses certain types of transitions and rather predicts a subset of those which are most likely to occur. Eventually, to measure if the sampler has the ability to provide a representative set of the correct ensemble of pathways is to measure statistical quantities, which will be the focus of future research.

\section{Acknowledgments}
This work was partially supported by the U.S. Department of Energy Advanced Scientific Computing Research (ASCR) under DOE-FOA-2493 ``Data-intensive scientific machine learning''.  DP acknowledges support by the U.S.\ Department of Energy, Office of Science, Office of Basic Energy Sciences, Heavy Element Chemistry Program under contract KC0302031 at Los Alamos National Laboratory (LANL).  Los Alamos National Laboratory is operated by Triad National Security, LLC, for the National Nuclear Security Administration of the U.S.\ Department of Energy (contract no.\ 89233218CNA000001). HT acknowledges funding provided by the LANL Director's office, and support by the Los Alamos Employees Scholarship Fund (LAESF). This work was supported in part by the U.S. Department of Energy, Office of Science, Office of Workforce Development for Teachers and Scientists (WDTS) under the Science Undergraduate Laboratory Internships (SULI) Program. This research used resources provided by the Darwin testbed at LANL which is funded by the Computational Systems and Software Environments subprogram of LANL's Advanced Simulation and Computing program (NNSA/DOE).

\bibliography{sources}

\end{document}